\documentclass[apj, iop, numberedappendix]{emulateapj}

\usepackage[utf8]{inputenc}
\usepackage[T1]{fontenc}
 
\usepackage[backref,breaklinks,colorlinks,citecolor=blue]{hyperref}

\usepackage{ulem}

\shorttitle{Constraints on the End of Reionization}
\shortauthors{Christenson et al.}

\usepackage{graphicx}
\usepackage{amssymb}
\usepackage{amsmath}
\usepackage{cleveref}

\usepackage{xcolor}
\usepackage{ulem}

\usepackage{natbib}
\usepackage{tabularx} 
\usepackage{multirow} 

\begin{document}

\newcommand{\kms}{{km\ s$^{-1}$}}
\newcommand{\flux}{ergs\ s$^{-1}$\ cm$^{-2}$}
\newcommand{\lum}{ergs\ s$^{-1}$}
\newcommand{\lya}{Ly$\alpha$}

\newcommand{\numLAESa}{{641}} 
\newcommand{\numLAESb}{{428}} 

\def\gdb#1{{\color{orange}{#1}\color{black}}}
\def\hmc#1{{\color{magenta}{#1}\color{black}}}
\def\cut#1{{\color{orange}\sout{#1}\color{black}}}


\title{Constraints on the End of Reionization from the Density Fields Surrounding Two Highly Opaque Quasar Sightlines}

\author{Holly M. Christenson\altaffilmark{1}, George D. Becker\altaffilmark{1}, Steven R. Furlanetto\altaffilmark{2}, Frederick B. Davies\altaffilmark{3}, Matthew A. Malkan\altaffilmark{2}, Yongda Zhu\altaffilmark{1}, Elisa Boera\altaffilmark{4}, Adam Trapp\altaffilmark{2} }

\altaffiltext{1}{Department of Physics  Astronomy, University of California, Riverside, CA 92521, USA; hchri004@ucr.edu}
\altaffiltext{2}{Department of Physics and Astronomy, University of California, Los Angeles, CA 90095, USA}
\altaffiltext{3}{Max Planck Institut f$\ddot{\rm{u}}$r Astronomie, Heidelberg, Germany}
\altaffiltext{4}{INAF — Osservatorio Astronomico di Trieste, Trieste, Italy}

\begin{abstract}
The observed large-scale scatter in \lya\ opacity of the intergalactic medium at $z<6$ implies large fluctuations in the neutral hydrogen fraction that are unexpected long after reionization has ended.
A number of models have emerged to explain these fluctuations that make testable predictions for the relationship between \lya\ opacity and density. We present selections of $z=5.7$ \lya-emitting galaxies (LAEs) in the fields surrounding two highly opaque quasar sightlines with long \lya\ troughs. The fields lie  towards the $z=6.0$ quasar ULAS J0148+0600, for which we re-analyze previously published results using improved photometric selection, and towards the $z=6.15$ quasar SDSS J1250+3130, for which results are presented here for the first time. In both fields, we report a deficit of LAEs within 20 $h^{-1}$ Mpc of the quasar. The association of highly opaque sightlines with galaxy underdensities in these two fields is consistent with models in which the scatter in \lya\ opacity is driven by large-scale fluctuations in the ionizing UV background, or by an ultra-late reionization that has not yet concluded at $z=5.7$. 

\end{abstract}

\keywords{Reionization, Galaxies: Intergalactic Medium - High Redshift, Quasars: Absorption Lines
}

\section{Introduction}\label{sec:intro}

Cosmic reionization was the last major phase transition in the history of the universe, during which radiation from the first luminous sources ionized neutral hydrogen in the intergalactic medium (IGM) and transitioned the universe from a mostly neutral to a highly ionized state (see \citealt{wise19} for a review). The physical properties of the IGM at reionization redshifts can be used to constrain the timing, duration, and sources of reionization, which have major implications on our understanding of the first luminous sources in the universe and their environments.

A number of observations now suggest that much of the IGM was reionized from $z\sim6-8$.  Measurements of the cosmic microwave background (CMB) are consistent with an instantaneous reionization occurring at $z\sim7.7\pm0.7$ \citep{planck2020}.  Evolution in the fraction of UV-selected galaxies that show \lya\ in emission suggests that significant portions of the universe remain neutral at $z\sim7-8$ \citep[][and references therein]{mason18, jung20, morales21}.  The presence of damping wings in $z\geq7$ quasar spectra \citep{mortlock11,greig17,banados18,davies18b,greig19,wang20} also suggest a largely neutral IGM at those redshifts.  Meanwhile, the onset of \lya\ transmission in quasar spectra suggests that reionization was largely complete by $z \sim 6$ \citep{fan06,mcgreer11,mcgreer15}. 

Recent studies, however, have suggested that signs of reionization may persist in the IGM considerably later than $z = 6$. Measurements of the Ly$\alpha$ forest towards high-redshift QSOs show a large scatter in the opacity of the IGM to \lya\ photons at redshifts $\leq6.0$ \citep{fan06,becker15,bosman18,eilers18,yang20,bosman21}, which is unexpected long after reionization has ended. The observed scatter on 50 comoving $h^{-1}$ Mpc scales has been shown to be inconsistent with simple models of the IGM that use a uniform ultraviolet background (UVB, \citealt{becker15,bosman18,eilers18,yang20,bosman21}) . The most striking example of this scatter is the large Gunn-Peterson trough associated with the $z=6.0$ quasar ULAS J0148+0600  (hereafter J0148), which spans 110 $h^{-1}$ Mpc and is centered at $z=5.7$ \citep{becker15}. While some scatter in \lya\ opacity is expected due to variations in the density field \citep[e.g.,][]{lidz06}, the extreme opacity in the J0148 field cannot be explained by variations in the density field alone. Several types of models have therefore emerged to explain the observed scatter as due to variations in the IGM temperature and/or ionizing background, or potentially the presence of large neutral islands persisting below redshift six.

One type of model is based on a fluctuating ultraviolet background, in which large-scale fluctuations in the photoionizing background drive the large-scale fluctuations in \lya\ opacity. Galaxy-driven UVB models, in which the fluctuations in the ionizing background result from clustered sources and a short, spatially variable mean free path, have been considered by \citet{davies16}, \citet{daloisio18}, and \citet{nasir20}.  In this scenario, highly opaque regions are associated with low-density voids that contain few sources and therefore have a suppressed ionizing background. Low-opacity regions, in contrast, would have a strong ionizing background from its association with an overdensity of galaxies. Alternatively, \citet{chardin15,chardin17} proposed a model in which the ionizing background is dominated by rare, bright sources such as quasars, which naturally produces spatial fluctuations in the UVB. Because quasars are rare, bright sources, the resulting UVB is not tightly coupled to the density field. In this scenario, a trough is associated with a suppressed ionizing background due to a lack of nearby quasars. The quasar-driven model, however, is somewhat disfavored because the required number density of quasars is at the upper limit of observational constraints and may also be in conflict with observational constraints on helium reionization \citep{daloisio17,mcgreer18,garaldi19}

\citet{daloisio15} proposed a model in which the opacity fluctuations are driven by large spatial variations in temperature, leftover from a patchy reionization process. In this scenario, overdense regions were among the first to reionize, and therefore have had more time to cool than less dense, more recently reionized regions. Absorption troughs such as the one towards J0148 are associated with overdense regions in this scenario; conversely, highly transmissive regions would be underdense.

More recently, a new type of model has emerged that suggests reionization may have ended later than $z\sim6$, as widely assumed \citep{kulkarni18,keating20a,keating20b,nasir20, choudhury21,qin21}. In this scenario, the observed scatter in \lya\ opacity is driven at least partly by islands of neutral hydrogen remaining in the IGM past $z=6$. Troughs like the one associated with J0148 therefore trace regions of the IGM that have not yet been reionized. The last places to become ionized in this model are low density, but those same underdense regions may quickly become highly transmissive once they have been reionized \citep{keating20b}.  These models predict that both high- and low-opacity sightlines may be underdense (although see \citealt{nasir20}, who find a large range in densities for transmissive lines of sight). We note that ultra-late reionization models typically also include a fluctuating UVB, but their defining feature is the presence of neutral islands at $z < 6$.

A key result of the attempts to model large-scale fluctuations in \lya\ opacity is that each type of model makes strong predictions for the relationship between opacity and density, particularly for extremely high and low opacities. Both of these quantities can readily be measured; the opacity of a sightline can be obtained from a background quasar's \lya\ forest, and a galaxy survey can be used to trace the underlying density. \citet{davies18} demonstrated that surveys of Lyman alpha emitters (LAEs) should be able to distinguish between these models for extremely high- and low-opacity sightlines. LAEs are a good choice for this type of observation because LAE surveys at $z \sim 6$ can be conducted with only three bands of photometry.  Narrow-band filters tuned to the atmospheric window near 8200~\AA, corresponding to \lya\ at $z = 5.7$, are also well matched to a redshift where large opacity fluctuations are present.

The results of a LAE survey in the J0148 field were published in \citet{becker18}. These results were consistent with fluctuating UVB and late reionization models, and strongly disfavored the fluctuating temperature model. \citet{kashino20} followed up with a selection of Lyman break galaxies in the same field as a separate probe of density, and also reported a strong underdensity associated with the trough.

In this paper, we extend the study of the \lya\ opacity-density relation to a second field surrounding a highly opaque quasar sightline.  We provide an updated selection of LAEs towards ULAS J0148+0600, and present new results for SDSS J1250+3130, whose spectrum exhibits an 81 comoving $h^{-1}$ Mpc \lya\ trough.  The LAE selections are based on updated LAE selection criteria, which we verify with spectroscopic followup of J0148 LAEs with Keck/DEIMOS. We summarize the observations in Section~\ref{sec:obs}, and describe the photometry and LAE selection criteria in Section~\ref{sec:methods}. The accompanying spectroscopy is presented in Appendix B.  We present the results of LAE selections in both fields in Section~\ref{sec:results}, and compare the results to current reionization models in Section~\ref{sec:analysis} before summarizing in Section~\ref{sec:summary}. Throughout this work, we assume a $\Lambda$CDM cosmology with $\Omega_m=0.3$, $\Omega_{\Lambda}=0.7$, and $\Omega_b=0.048$. All distances are given in comoving units, and all magnitudes are in the AB system.

\section{Observations}\label{sec:obs}
\begin{figure*}
\includegraphics[width=\textwidth]{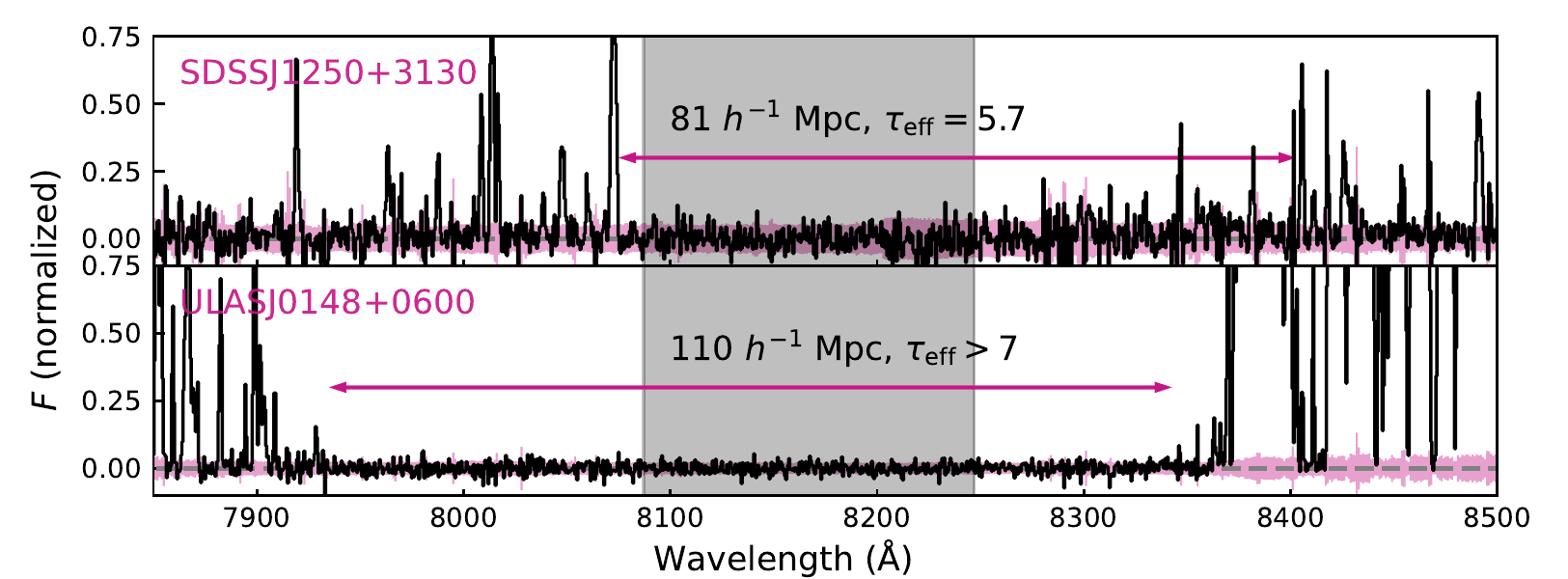}
\caption{Partial spectra of the two quasars whose fields we observe with Subaru/HSC. The top panel shows a Keck/ESI spectrum of SDSS J1250+3130, which exhibits a \lya\ trough that is 81 $h^{-1}$ Mpc in length with $\tau_{\rm eff} = 5.7 \pm 0.4$} (Zhu et al, in prep.).  The bottom panel shows an X-Shooter spectrum of ULAS J0148+0600, which exhibits a 110 $h^{-1}$ Mpc \lya\ trough with $\tau_{\rm eff}\geq7$ \citep{becker15}. The approximate extent of each trough is indicated by the pink arrows. These quasars represent some of the most extreme \lya\ troughs known at $z < 6$. The shaded gray region shows wavelengths covered by the NB816 filter with at least 10\% transmittance, which corresponds to \lya\ at $z \simeq 5.7$. The shaded pink region indicates the $\pm1\sigma$ uncertainty interval.
\label{fig:qso_spectra}
\end{figure*}

Imaging data taken with Subaru Hyper Suprime Cam (HSC) were previously presented for the ULAS J0148+0600 field by \citet{becker18}. The spectrum of ULAS J0148+0600 contains a 110 $h^{-1}$ Mpc trough that has effective optical depth of $\tau_{\rm eff} \geq7$, where $\tau_{\rm eff}=-\rm{ln}\langle$ $F$ $\rangle$ and $F$ is the mean continuum-normalized flux. For this work, we obtained HSC imaging of a second field, towards the $z = 6.15$ quasar SDSS J1250+3130 (hereafter J1250). The \lya\ forest in the spectrum of J1250 contains a trough spanning 81 $h^{-1}$ Mpc with $\tau_{\rm eff} = 5.7 \pm 0.4$ (Zhu et al, in prep.).  The J1250 and J0148 fields represent some of the most highly opaque sightlines known at these redshifts. Figure~\ref{fig:qso_spectra} shows subsets of the X-Shooter spectrum for ULAS J0148+0600 \citep{becker15} and the Keck/ESI spectrum for SDSS J1250+3130, displaying their \lya\ troughs.

\LongTables
\begin{center}
\begin{deluxetable}{lcccccc}
\tablewidth{1.75\textwidth}
\tablecaption{Summary of HSC imaging \label{tab:obs_summary}}
\tablehead{\colhead{} & \colhead{Filter} & \colhead{$t_{exp}$ (hrs)} &
\colhead{ Seeing$^b$} & \colhead{$m_{5\sigma,PSF}^{c}$} & \colhead{$m_{5\sigma,1.5\arcsec}^c$} }
\startdata
\multirow{3}{*}{J0148}&$r2$ & 1.5 & 0.76 & 26.4 & 26.0 \\
&$i2$ & 2.4 & 0.80 & 26.0 & 25.6\\
&$NB816$ & 4.5 & 0.73 & 25.2 & 25.0 \\
\hline
\multirow{3}{*}{J1250}&$r2$ & 2.0$^a$ & 0.83 & 26.4 & 26.2 \\
&$i2$ & 2.5 & 0.81 & 26.1 & 25.8 \\ 
&$NB816$ & 2.8 & 0.74 & 25.3 & 25.0
\enddata
\tablenotetext{a}{Partially observed in gray time.}
\tablenotetext{b}{Median seeing FWHM in combined mosaic.}
\tablenotetext{c}{Magnitude at which 50\% of detected sources have $S/N \geq 5$.}

\end{deluxetable}
\end{center}

~\label{obs_summary}

The J1250 field was observed via the HSC queue in April and June 2019, with the majority of the data being taken during dark time in April. Additional observations were taken during dark time in May 2020 and January 2021.  As for the J0148 field, we obtained imaging centered on the quasar position in the NB816 filter, which has a mean transmission-averaged wavelength $\lambda = 8177$ \AA, corresponding to Ly$\alpha$ emission at $z=5.728$, and two broadband filters, $i2$ and $r2$. The narrowband observations were completed as planned, but the initial $r2$ observations in the J1250 field were completed in gray time and supplemented by additional dark time observations in May 2020. We summarize the observations in both fields in Table~\ref{tab:obs_summary}.

We reduced the raw data with the LSST Science Pipeline, Versions 19 (J0148 field) and 22 (J1250 field) \citep{ivezic08,juric15}. The pipeline combines individual CCDs into stacked mosaics, using PanStarrs DR1 imaging \citep{chambers16} for astrometric and photometric calibrations. We use Source Extractor \citep{bertin96} to identify the spatial coordinates of sources in the final stacked mosaics, and then make photometric measurements at those positions based on PSF fitting, which we describe in more detail in Section~\ref{sec:methods}.

Table~\ref{tab:obs_summary} shows the median 5$\sigma$ limiting PSF and aperture magnitudes in each band for both fields. These values represent the magnitudes at which at least 50\% of the detected sources are measured at signal-to-noise ratios $S/N \geq 5$.

We also use the imaging data to independently measure the \lya\ opacity over the NB816 wavelengths along each quasar line of sight.  The results are presented in Appendix A.


\section{Methods}\label{sec:methods}
In this section we describe in detail the methods used to make photometric measurements and select LAE candidates.

\subsection{Photometry}\label{sec:psf_fitting}

\citet{becker18} used CModel fluxes generated by the LSST pipeline, which are a composite of the best-fit exponential and de Vaucouleurs profiles \citep{sdss2004,bosch18}. We verified the quality of the flux calibration by checking the fluxes of 25 objects in each field from the SDSS catalogs. While the flux measurements for the verification objects were accurate to within the photometric errors, fluxes for faint, typically seeing-limited objects were found to be less reliable. For some of these objects, the best-fit CModel profile resulted in conspicuously high fluxes that were not in agreement with the fixed-aperture and PSF fluxes. This systematic overestimation of CModel fluxes compromised the initial selection of LAEs in the J0148 field in two ways: objects that are not credible LAEs were selected as LAEs based on artificially high narrowband flux, and objects that could be credible LAEs were rejected based on artificially high broadband fluxes that resulted in failure of one or more color criteria. Examples of both types are shown in Appendix C.

To address these problems with the CModel fluxes, we implemented PSF measurements to replace the CModel measurements as the primary flux used in the analysis. The PSF photometry is optimized to maximize the detection of faint, often unresolved sources for the purposes of constructing a density map. Sources whose profiles are not well-represented by a PSF profile, such as extended sources, are assigned an aperture flux as their primary flux measurement, which we describe in more detail below. 
The photometry has the following steps:

\begin{enumerate}
\item At each source position identified and measured by Source Extractor in the combined mosaics, we measure the flux in a 1.5\arcsec\ aperture.
\item We then measure the median sky background measured in a 5\arcsec\ annulus around the aperture, excluding any pixels that are flagged by the data reduction pipeline as sources.
\item A 2-dimensional Gaussian profile is fit over a stamp of the combined mosaic $10\times10$\arcsec\ in size centered on the source, using the measured sky background as the offset and holding the FWHM fixed to the median seeing.  The only parameter allowed to vary is the amplitude.
\item Each pixel in the stamp is compared to the resulting fit. Pixels that differ from the model by more than five times the noise in the sky background are excluded from the next iteration of fitting. The primary purpose of this step is to reject cosmic rays and bad pixels.
\item The 2D Gaussian is fit again, excluding outlier pixels. After re-fitting, all pixels are again compared to the model and the exclusion and re-fitting process is repeated. Pixels that were previously rejected may be included in the next iteration of the fit. If the fitting exceeds ten iterations, more than 5\% of pixels in the stamp are rejected, or more than 5\% of the pixels within a 1.5\arcsec\ aperture are rejected, the fit is considered a failure and the aperture flux is used as the primary flux measurement for that object. Typically, extended sources and other objects whose profiles are not well represented by the PSF profile will therefore be assigned aperture fluxes. If re-fitting fails to improve the fit (the same set of pixels are selected for exclusion in two subsequent iterations) but the maximum number of iterations and excluded pixels are not exceeded, the fit is considered a success and the resulting PSF flux is recorded. Approximately 20\% of all sources fail, and 50\% are refit at least once, most undergoing two iterations. 
\end{enumerate}

This PSF measurement is conducted for each band, independently of the others. We have allowed the fitting routine to default to aperture fluxes because for many credible LAEs, the $r2$ and $i2$ fluxes are formally undetected, and the results of fitting a Gaussian to a field dominated by noise may be unpredictable. In these cases, we default to the aperture flux rather than accept a potentially bad fit.

\subsection{LAE selection procedure}\label{sec:lae_selection}
In addition to improving our photometric measurements, we have adjusted the criteria we use for selecting LAE candidates. Our observations in the J1250 field were made over the course of three years, and the partially complete observations had large variations in depth across the three bands. This disparity motivated an adjustment of the selection criteria to account for the depth in each band.  Our completed observations are still slightly uneven in depth across the three filters, and there are variations in depth between fields - for example, the J1250 field is slightly deeper than the J0148 field in both broadband filters. The revised selection criteria described do not dramatically change the LAE selection in these two fields; however, they reduce the number of selected objects by $\sim25\%$. We emphasize that all sources must still pass a visual inspection to be accepted as LAE candidates. 

The criteria originally used to select LAEs in the J0148 field were based on those used in \citep{ouchi08}: $NB816\leq26.0$, $S/N_{NB816} \geq 5$, $i2-NB816\geq1.2$, and $r2 \leq 2\sigma_{r2}$ \textbf{or} $r2 \geq 2\sigma_{r2}$ \textbf{and} $r2-i2 \geq 1.0$. These requirements are designed to select line emitters and rule out low-redshift objects, but have no requirement for uncertainty or $S/N$ in any band except the narrow band. 

In order to account for the different depths of our photometric bands, we re-express the selection criteria in terms of probability densities. For each color cut, we require that at least 50\% of the probability density for that color is above the minimum acceptable color. We also require that 95\% of the probability density be greater than the 1$\sigma$ lower limit for an object with $S/N_{NB}=5$ and $i2-NB816=1.2$. The second requirement is designed to exclude objects that meet the minimum $i2-NB816$ requirement but with large uncertainties.


Calculating the probability density for the color of each object is complicated somewhat by fluxes that are formally undetected. 
To calculate a physically motivated uncertainty for a color that is based on a non-detected flux (which may be negative), we used a set of artificial sources to generate probability density functions (PDF) for non-detected fluxes, with the prior that the true flux must be positive. We added artificial sources with known, positive fluxes ($F_{true}$) in random positions across the field and then measured the PSF fluxes ($F_{meas}$) of these artificial sources as previously described. The distribution of $F_{true}$ values associated with objects that have a given $F_{meas}$ represents a PDF that can be used for assessing the uncertainty in an object’s color. The resulting PDF is a Gaussian centered on $F_{meas}$ and FWHM$\sim\sigma_{meas}$, with negative values truncated. We therefore take the probability density function for measured flux values associated with real sources, positive and negative, to be a Gaussian with $\mu=F_x$ and $\sigma=\sigma_{x}$, with negative values truncated and re-normalized to unity.  


For simplicity, we express the color criteria as flux ratios. To find the PDF of a flux ratio, we first generate a PDF for each flux value as described above. We then take the ratio of each possible combination of values from the one-dimensional PDFs to generate a two-dimensional PDF for the flux ratio. We then find the total probability that the flux ratio exceeds the minimum color threshold to evaluate the selection criteria.

In addition to the color cuts described above, we also require that $F_{NB816} \geq 7.6 F_{r2}$ (or, $r2-NB816 \geq 2.2$) with at least a 50\% probability. This requirement follows from the $i2 -NB816$ and $r2-i2$ colors above, and is expected due to the decreasing transmission of blue flux from high-redshift objects. This additional check helps to exclude objects with a significant probability of being low-redshift contaminants. 

Finally, we adopted a narrow band limit of $NB \leq 25.5$. This is somewhat brighter than the limit of $NB816 \leq 26.0$ used by \citet{becker18}. The brighter limit was chosen because, after making completeness corrections (see Section~\ref{sec:results}), we found that our observations were only $\sim10\%$ complete in the $25.5\leq NB816\leq26.0$ bin. We selected an additional 143 objects in this bin, although they are excluded from the analysis because of the poor completeness.


To summarize, the final selection criteria applied to our LAEs are as follows:

\begin{itemize}
\item $NB \leq 25.5$
\item $S/N_{NB816} \geq 5$
\item $\frac{F_{NB816}}{F_{i2}} \geq 3.0$ (50\% probability) and $\frac{F_{NB816}}{F_{i2}} \geq 1.7$ (95\% probability)
\item $F_{r2} \leq 2\sigma_{r2}$, or $F_{r2}\geq2\sigma_{r2}$ and $F_{i2}/F_{r2}\geq 2.5$ 
\item $\frac{F_{NB816}}{F_{r2}} \geq 7.6$ (50\% probability) and $\frac{F_{NB816}}{F_{r2}} \geq 4.0$ (95\% probability)
\end{itemize}


Finally, objects that pass these criteria are inspected visually to remove moving or spurious sources.

To summarize, our selection criteria are based on ones used in previous works to detect LAEs at $z=5.7$ with Subaru \citep[e.g.,][]{ouchi08,konno18,shibuya18a,ouchi18}, but with some modifications.
The main differences are that we impose additional probability requirements for the color criteria and add a $\frac{F_{NB816}}{F_{r2}}$ requirement.  Following \citet{diaz14}, we also do not make use of a bluer filter to exclude low-redshift contaminants.
We do, however, use a more selective $r2$ criterion than \citet{ouchi18} and \citet{shibuya18a}, who require that LAEs are undetected in $r2$ at $3\sigma$ (compared to $2\sigma$ in this work) unless they satisfy the $r2-i2$ color cut.

Spectroscopic follow-up of a subset of LAEs in the J0148 field with Keck/DEIMOS suggests that our selection criteria should yield a high-quality sample of LAEs.  We present details of the spectroscopy in Appendix B.


\section{Results}\label{sec:results}
\begin{figure}
\includegraphics[width=0.45\textwidth]{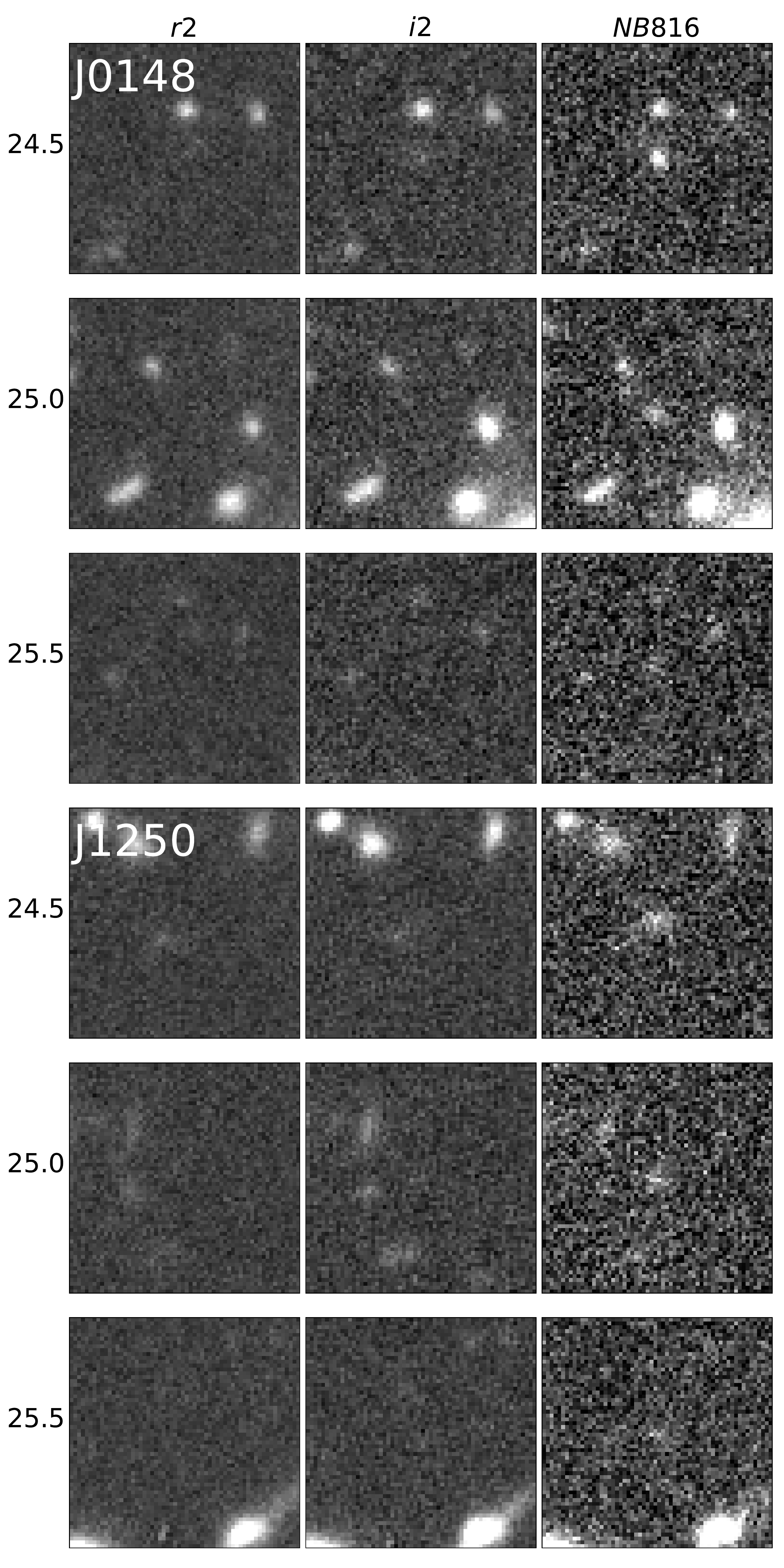}
\caption{Example LAE candidates selected in the J0148 (top three rows) and J1250 (bottom three rows) fields with the criteria described in Section ~\ref{sec:lae_selection}. The cutout images are 10\arcsec\ on each side and centered on the LAE position. Each row shows images of a sample candidate of a different narrowband magnitude (shown at the left) in the $r2$, $i2$, and $NB816$ bands (left to right). The sample candidates were chosen to have $S/N_{NB816}$ values near the median for objects at similar NB816 magnitudes.}
\label{fig:lae_stamps}
\end{figure}

We now turn to the results of the photometric selection. Using the  procedure outlined in Section~\ref{sec:lae_selection}, we select \numLAESa\ LAEs in the J0148 field and \numLAESb\ LAEs in the J1250 field. The number of LAEs selected in the J0148 field is somewhat lower than found by \citet{becker18}.  We discuss the reasons for this difference in Appendix C, but note that the overall spatial distribution of sources is similar. Cutout images for example LAE candidates selected in the J0148 field (top three rows) and J1250 field (bottom three rows) are shown in Figure~\ref{fig:lae_stamps}. The cutout images are 10\arcsec\ on each side and centered on the LAE candidate. Each row shows an example candidate of a different narrowband magnitude (shown at the left) in the $r2$, $i2$, and $NB816$ bands (left to right). The examples were chosen to have $S/N_{NB816}$ near the median value for objects of similar magnitude. 

\begin{figure*}
\includegraphics[width=\textwidth]{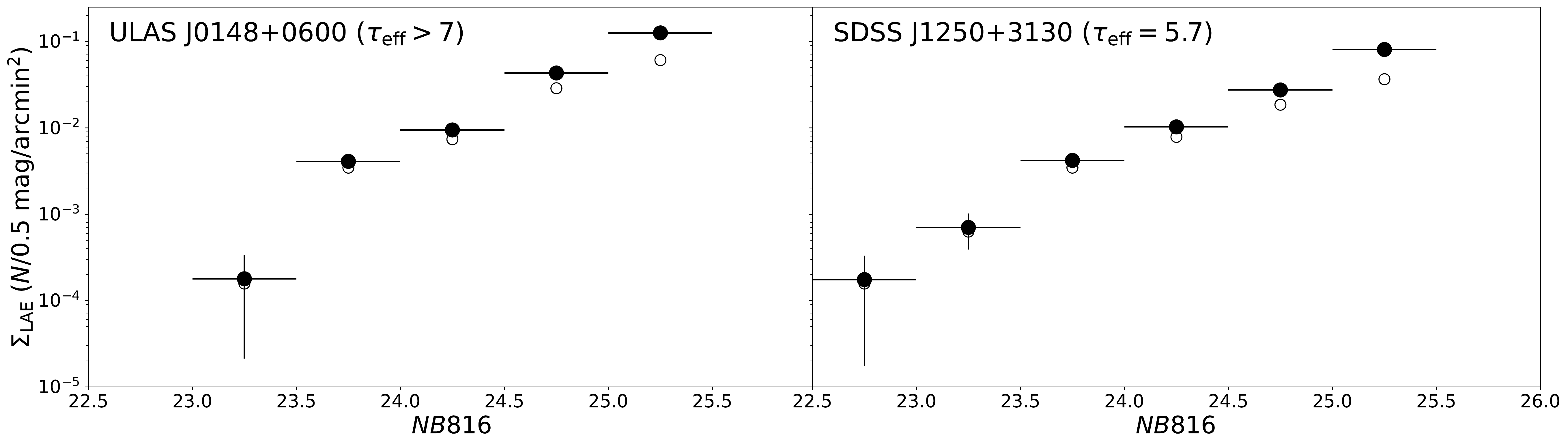}
\caption{Surface density of LAE candidates in the J0148 (left) and J1250 (right) fields. The LAE candidates are binned in 0.5 mag increments. Raw surface densities are shown with open markers. Filled markers show the surface densities with a correction made for completeness (see Section~\ref{sec:results} and Appendix D}). The error bars on the corrected measurements are 68\% Poisson intervals.
\label{fig:sd_mag}
\end{figure*}

The surface density of the LAE candidates within 45\arcmin\ of the quasar position in both fields as a function of their $NB816$ magnitude is shown in Figure~\ref{fig:sd_mag}. Raw values are shown with open markers, and completeness-corrected values are shown with filled markers. We calculate the completeness correction as a function of both distance from the quasar position and NB816 magnitude by injecting a catalog of artificial LAE candidates across the field, then putting them through the LAE selection procedure. The completeness correction applied to the real LAE candidates is then given by the reciprocal of the fraction of artificial LAEs detected in each bin. The correction factor adjusts for variations in sensitivity across the field and for loss of area covered by bright foreground sources. The completeness as a function of NB816 magnitude and distance from the quasar for both fields is given in Appendix D.

\begin{figure*}
\includegraphics[width=\textwidth]{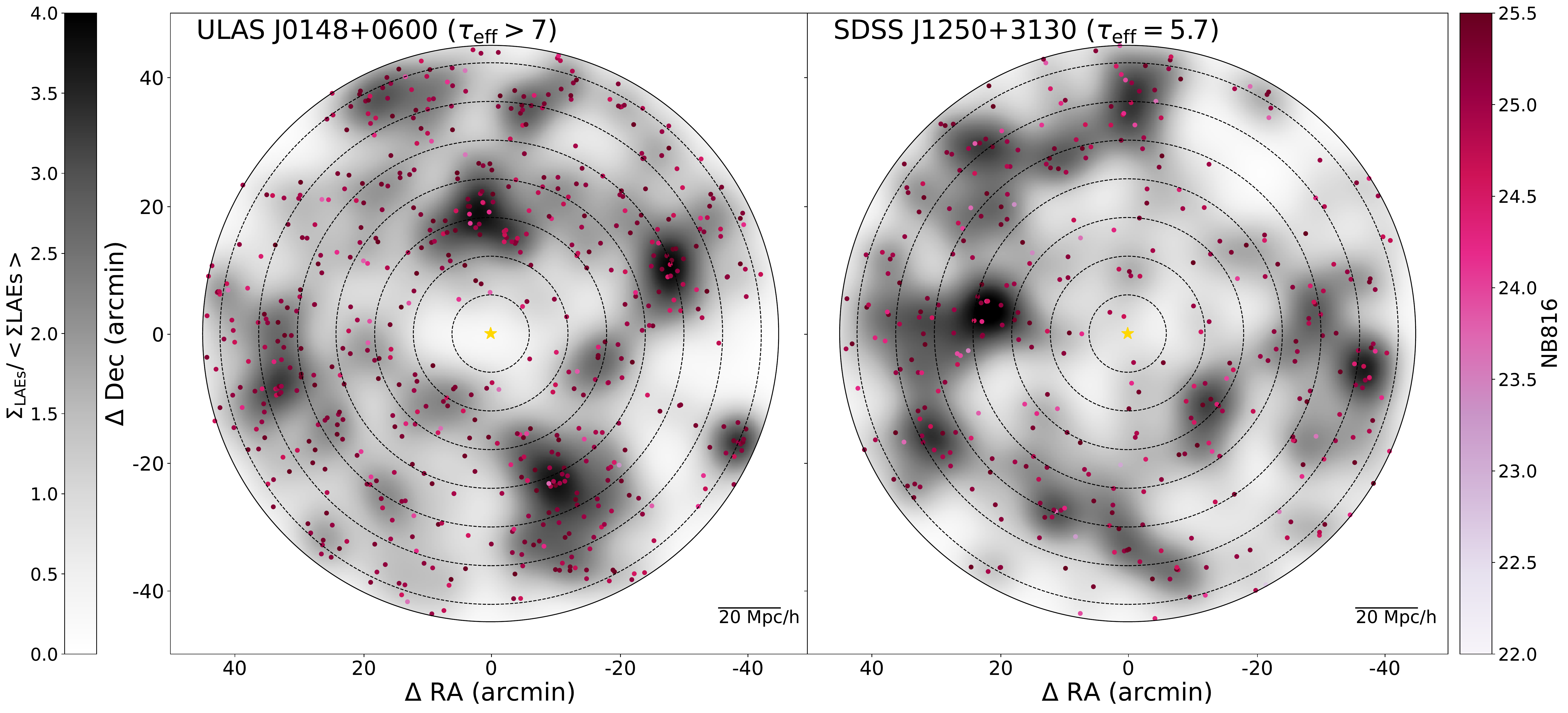}
\caption{Distribution of LAE candidates in the J0148 (left) and the J1250 (right) fields. Each field is shown centered on the quasar (gold star). LAE candidates are shown with a color that indicates their narrowband magnitude. Shading indicates the surface density of LAE candidates at each position, which is is calculated by kernel density estimation using a Gaussian kernel and normalized by the mean surface density measured across the entire field. Concentric, dotted circles are shown in increments of 10 $h^{-1}$ Mpc  projected distance from the quasar. The solid circle marks the edge of the field of view, 45\arcmin\ from the quasar. 
}
\label{fig:lae_sd_map}
\end{figure*}

The spatial distribution of selected LAEs in each field is shown in Figure~\ref{fig:lae_sd_map}. LAE candidates are shown with colors corresponding to their NB816 magnitudes. The quasar is centered in each field and denoted with a star.  Dotted concentric circles are plotted in increments of 10 $h^{-1}$ Mpc. The solid outer circle shows the edge of the field of view, 45\arcmin\ from the quasar. 

LAE candidates are shown plotted over a surface density map.  We create the surface density map for the LAE candidates in each field by superimposing a regular grid of 0.24\arcmin\ (0.4 $h^{-1}$ Mpc ) pixels onto the field. In each grid cell, we find the surface density by kernel density estimation using a Gaussian kernel of bandwidth 1.6 arcmin. We then normalize the grid by the average surface density of the field.

\begin{figure*}
\includegraphics[width=\textwidth]{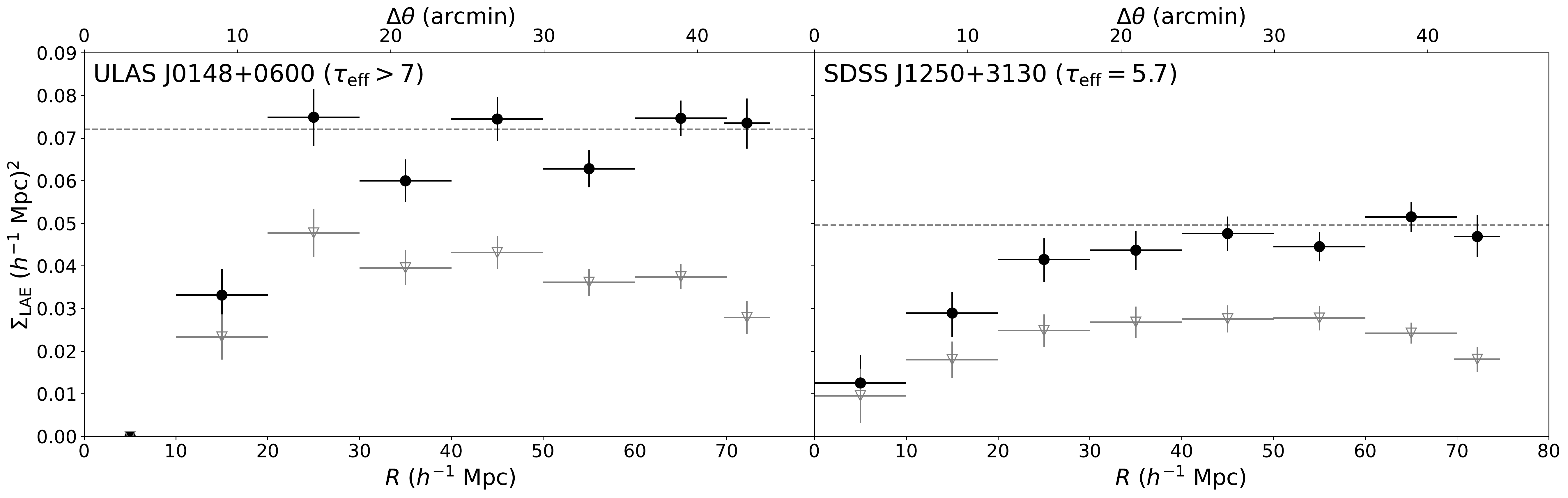}
\caption{Surface density of LAEs in the J0148 (left) and J1250 (right) fields. The filled black circles show corrected surface density, and the unfilled gray triangles show the uncorrected measurements.} The surface density is measured as a function of projected distance from the quasar in annual bins of 10 $h^{-1}$ Mpc, except for the outermost bin which is 4.5 $h^{-1}$ Mpc. 
The dotted line represents the mean surface density of LAE candidates that lie within $15\arcmin \leq \Delta \theta \leq 40\arcmin$ of the quasar. Horizontal error bars show the width of the annuli, and vertical error bars are 68\% Poisson intervals.
\label{fig:rad_combined}
\end{figure*}

We calculate the surface density of the LAE candidates as a function of radius. The raw surface density is measured in 10 $h^{-1}$ Mpc concentric annuli centered on the quasar position and then corrected for completeness. The corrected surface density is shown as a function of projected distance from the quasar for each field in Figure~\ref{fig:rad_combined}. The horizontal line represents the mean background surface density of LAE candidates, averaged over $15\arcmin \leq \Delta\theta \leq 40\arcmin$. The surface density measurements for the J0148 and J1250 fields are summarized in Tables ~\ref{tab:num_dens_j0148} and ~\ref{tab:num_dens_j1250}, respectively.

The key result from \citet{becker18} is unchanged; Figure~\ref{fig:lae_sd_map}
shows a marked underdensity within 20 $h^{-1}$ Mpc of the quasar in the J0148 field. The LAE catalog presented here and that presented in \citet{becker18} are largely consistent within the expected variations in LAE selection at the faintest magnitudes, where the sample is $\sim50$\% complete, and display the same large-scale structures. We estimate that 15\% of the objects appearing in each catalog are affected by the flux issues discussed in Section~\ref{sec:psf_fitting}. A more detailed comparison is given in Appendix C.

We also find a deficit of LAEs in the inner 20 $h^{-1}$ Mpc of the J1250 field. This result is consistent with the J0148 field, and confirms the association between highly opaque sightlines and underdense regions in a second field.  

We note that the two fields vary in the observed surface density of LAEs; we select \numLAESa\ LAEs in the J0148 field and \numLAESb\ in the J1250 field in the same survey volume.  While our main result is based on a differential measurement of the LAE surface density \textit{within} each field, one might also wonder about the variance in LAE density \textit{between} the two fields. We can gauge whether this variance is reasonable using a simple linear bias treatment, which is accurate for the large volume probed by our survey (see, e.g., \citealt{trapp20}). Using the \citet{trac15} halo mass function and its linear bias expansion with the standard scaling method \citep{tramonte17, trapp20}, we expect $\sim 535 \pm 100$ halos of mass $\sim 1.7 \times 10^{11} \ M_\odot$ dark matter halos in each of our fields, where the ``error” is the $1\sigma$ sample variance. In this scenario, the two fields are within $\sim 1 \sigma$ of the expected value. If, however, only one quarter of halos contain LAEs, the number density would correspond to $7 \times 10^{10} \ M_\odot$ halos, which have a fractional standard deviation due to sample variance of $\sim 0.16$, still consistent with the observed fields. Both of these scenarios are reasonable in light of independent measurements of LAE properties at $z \sim 6$.  For example, \citet{khostovan19} estimate halo masses $\sim 10^{11} \ M_\odot$ for LAEs via clustering, while \citet{stark10} find that $\sim$25--50\% of galaxies have strong Lyman-$\alpha$ emission lines.  

{\citet{gangolli20} similarly find that large-scale structure is sufficient to explain the significant field-to-field variations of $z = 5.7$ LAEs in the SILVERRUSH survey \citep{ouchi18}. In contrast, they argue that patchy reionization is unlikely to drive these variations because, at the end of reionization, the neutral gas is largely confined to voids, where it should obscure fewer galaxies.  We note that our fields are somewhat unusual in that they were selected to have high IGM \lya\ opacities at the field center.  Even so, the overall variation in number mean density between fields appears to be consistent with cosmic variance in the number density of LAE hosts at this redshift.}

\LongTables
\begin{deluxetable}{lccc}
\tablewidth{0pc}
\tablecaption{LAE Number Density in the J0148 Field \label{tab:num_dens_j0148}}
\tabletypesize{\scriptsize}
\tablehead{\colhead{R (Mpc)}  & \colhead{$N_{LAEs}$} & \colhead{$N_{corr}^a$} & \colhead{$\Sigma$ LAE (Mpc h$^{-1}$)$^2$}}
\startdata
$5 (0-10)$  & $0$ & 0 & 0.00 \\
$15 (15-25)$  & $22$ & 31 & 0.033 ($0.028-0.039$)\\
$25 (20-30)$  & $75$ & 118 & 0.075 ($0.068-0.081$)\\
$35 (30-40)$  & $87$ & 132 & 0.060 ($0.055-0.065$) \\
$45 (40-50)$  & $122$ & 211 & 0.075 ($0.069-0.080$) \\
$55 (50-60)$  & $125$ & 217 & 0.063 ($0.058-0.067$)\\
$65 (60-70)$  & $153$ & 305 & 0.075 ($0.071-0.079$) \\
$72 (70-74.5)$  & $57$ & 150 & 0.074 ($0.68-0.079$) 
\enddata
\tablenotetext{a}{Completeness corrected}
\end{deluxetable}~\label{num_dens_j0148}

\LongTables
\begin{deluxetable}{lccc}
\tablewidth{0pc}
\tablecaption{LAE Number Density in the J1250 Field \label{tab:num_dens_j1250}}
\tabletypesize{\scriptsize}
\tablehead{\colhead{R (Mpc)}  & \colhead{$N_{LAEs}$} & \colhead{$N_{corr}^a$} & \colhead{$\Sigma$ LAE (Mpc h$^{-1}$)$^2$}}
\startdata
$5 (0-10)$  & 3 & 4 & 0.013 ($0.006-0.019$) \\
$15 (15-25)$  & 17 & 27 & 0.030 ($0.023-0.034$) \\
$25 (20-30)$  & 39 & 65 & 0.042 ($0.036-0.046$)\\
$35 (30-40)$  & 59 & 96 & 0.044 ($0.039-0.048$) \\
$45 (40-50)$  & 78 & 135 & 0.048 ($0.043-0.052$) \\
$55 (50-60)$  & 96 & 154 & 0.045 ($0.041-0.048$)\\
$65 (60-70)$  & 99 & 210 & 0.052 ($0.048-0.055$) \\
$72 (70-74.5)$  & 37 & 196 & 0.047 ($0.042-0.052$)
\enddata
\tablenotetext{a}{Completeness corrected}
\end{deluxetable}
~\label{num_dens_j1250}

\section{Analysis}\label{sec:analysis}
\subsection{Comparison to models for opaque sightlines}\label{sec:opaque_models}

\begin{figure*}[h!]
\includegraphics[width=\textwidth]{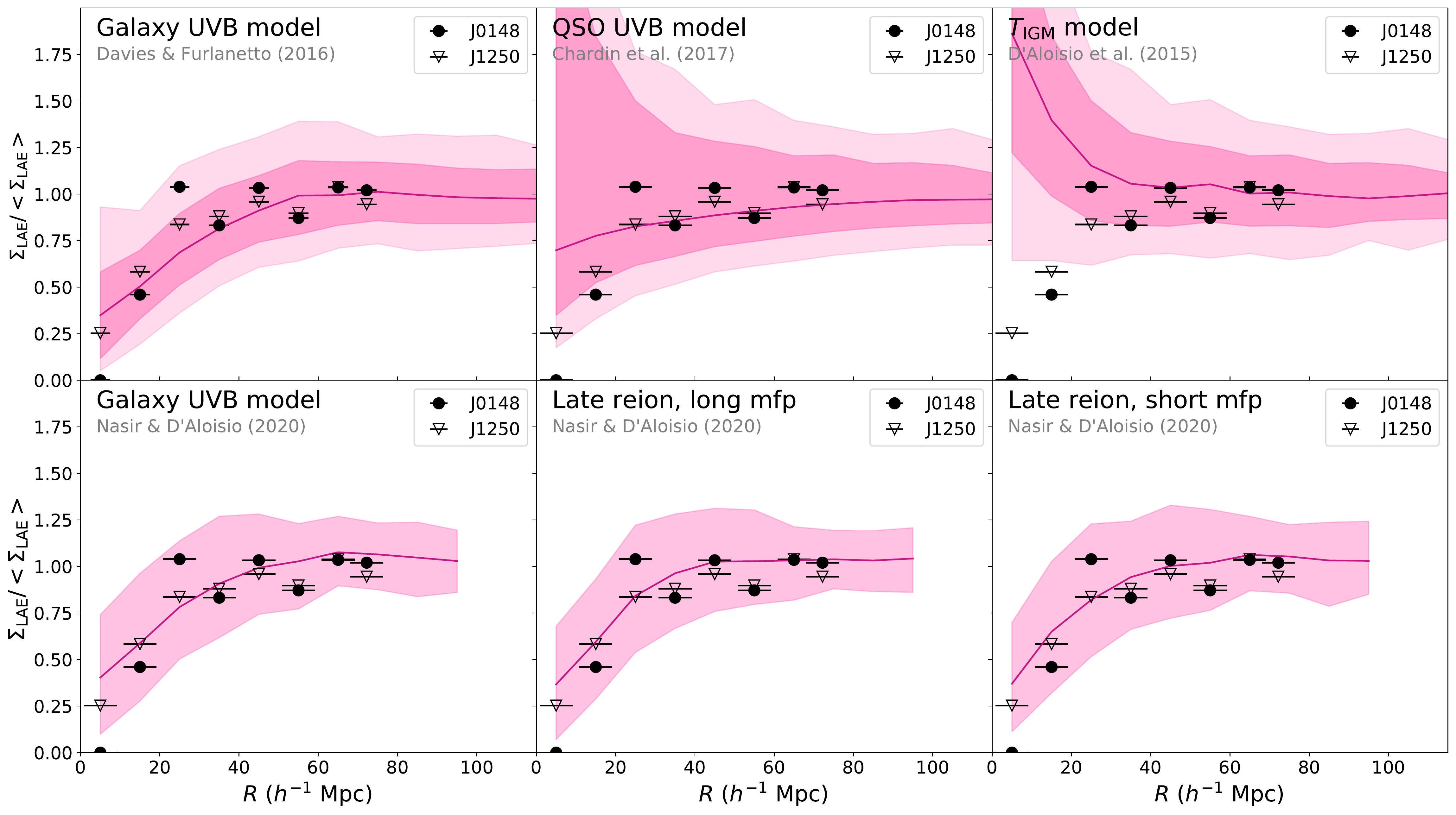}
\caption{Comparison of the observed radial distribution of LAE candidates in the J0148 (filled circles) and J1250 (open triangles) fields to model predictions.  The top row shows predictions from the galaxy UVB model based on \citet{davies16} (top left), the QSO UVB model based on \citet{chardin15,chardin17} (top center), and the fluctuating temperature model from \citet{daloisio15} (top right).   The bottom row shows predictions from \citet{nasir20}, including their galaxy UVB (early reionization) model (bottom left), late reionization model with a long mean free path (bottom center), and late reionization model with a short mean free path (bottom right)}. All model predictions are for highly opaque lines of sight ($\tau_{\rm eff}\geq7$). The horizontal error bars on the measured data points indicate the width of the bins, which is 10 $h^{-1}$ Mpc for all except the outermost bin, which is 4.5 $h^{-1}$ Mpc. The solid lines show median model predictions, which are averaged over 10 $h^{-1}$ Mpc bins throughout the simulation. In the top panels, the dark- and light-shaded regions show 68\% and 95\% ranges subtended by the mock samples drawn from the simulation.  The shaded regions in the lower panels show the 90\% range. All surface densities are given as a fraction of the mean surface density measured over 15\arcmin\ $\leq\theta\leq $ 40\arcmin.
\label{fig:models_opaque}
\end{figure*}

We now compare our observations to models that attempt to explain the large-scale fluctuations in IGM \lya\ opacity at $z \lesssim 6$. We consider six variations on three main types of models: fluctuating UVB, fluctuating temperature, and ultra-late reionization. We refer the reader to the introduction for a more detailed description of these models.

The first type of model is defined by large-scale fluctuations in the UVB. We consider two galaxy-driven models, one from \citet{davies16} and one from \citet{nasir20}. In these models, UVB fluctuations are driven by the clustering of ionizing sources and a short and spatially variable mean free path. The \citet{nasir20} model is based on an ``early'' (completed by $z \simeq 6$) reionization simulation and also includes temperature fluctuations.  In these models, high-opacity lines of sight are typically associated with underdense regions, where the UVB is suppressed.  We also consider a quasar-driven UVB model based on \citet{chardin15,chardin17}.  In this model, high-opacity lines of sight may be associated with a wide range of densities provided that they are in regions far from quasars, where the UVB is low.  We note that this model is disfavored by the fact that quasars may only provide a small fraction of the UVB at these redshifts \citep[e.g.,][]{mcgreer18,parsa18,kulkarni19}, but consider it here for completeness. 

The second type of model is from \citet{daloisio15}, and is defined by large-scale temperature fluctuations. In this model, highly opaque lines of sight are associated with overdensities that reionized early and have had sufficient time to cool.  

The third type of model is defined by reionization being incomplete at $z=6$. We consider two ultra-late reionization models from \citet{nasir20}.  These models include both regions of neutral hydrogen and a fluctuating ionizing background driven by clustered ionizing sources and a finite mean free path.  At $z = 5.8$, the "long mean free path" model has a hydrogen neutral fraction of $\langle \chi_{\rm H\,I} \rangle = 0.14$ and a mean free path of $\langle \lambda_{\rm mfp}^{912} \rangle = 27~h^{-1}$ Mpc, while the "short mean free path" model has $\langle \chi_{\rm H\,I} \rangle = 0.10$ and $\langle \lambda_{\rm mfp}^{912} \rangle = 9~h^{-1}$ Mpc. 
Predictions for the late reionization models in \citet{nasir20} are qualitatively consistent with those from \citet{keating20b} for opaque lines of sight.

Predictions for the \citet{nasir20} models are taken directly from that work.  All others are as implemented in \citet{becker18}. The LAE modelling is done using a similar approach in all cases.  We refer the reader to these papers for details, but briefly summarize the method here.  Galaxies are assigned to dark matter halos via abundance matching to the measured UV luminosity function of \citet{bouwens15}. The  spectra are modeled with a power-law continuum and a \lya\ emission line, with rest-frame equivalent widths drawn from the empirically calibrated models of \citet{dijkstra12}.
The modelled LAE populations are then used to construct expected surface density profiles for highly opaque lines of sight. \citet{becker18} use sightlines with $\tau_{\rm eff}\geq7$ measured on 50 $h^{-1}$ Mpc scales. This scale is somewhat shorter than the lengths of the J0148 and J1250 troughs (110 $h^{-1}$ Mpc and 81 $h^{-1}$ Mpc respectively); however, \citet{davies18} compared predictions for the surface density of LAEs as a function of $\tau_{\rm eff}$ on both 50 $h^{-1}$ and 110 $h^{-1}$ Mpc scales and found that the predictions were not highly sensitive to this choice. \citet{nasir20} use the 100 longest troughs in each simulation to make their predictions, typically 80$-$100 $h^{-1}$ Mpc in length, which is comparable to the lengths of the J0148 and J1250 troughs.

We compare these model predictions to the measured LAE surface density in the J0148 and J1250 fields in Figure~\ref{fig:models_opaque}. The top panel shows, from left to right, the galaxy UVB model based on \citep{davies16}, the QSO UVB model based on \citep{chardin17}, and the fluctuating temperature model based on \citep{daloisio15}.  The lower panel shows the three models from \citet{nasir20}: the first (left) is a galaxy UVB model, the second (center) is the ultra-late reionization scenario with a long mean free path, and the third (right) is the ultra-late reionization scenario with a short mean free path. 

The predictions from each model are averaged over 10 $h^{-1}$ Mpc bins.   The solid lines show the median prediction. In the top panels, the dark- and light-shaded regions indicate the 68\% and 95\% ranges, respectively, subtended by the mock samples drawn from the simulation. In the lower panel, the shaded regions indicate the 10th and 90th percentiles. All surface densities are normalized by the mean surface density in the field measured over $15\leq \theta \leq 40$.  We note that these model predictions are made for sightlines with $\tau_{\rm eff}\geq7$, while the J1250 sightline has $\tau_{\rm eff} \simeq 6$. \citet{davies18}, however, find that the predictions for these opacities are very similar.


In both the J0148 and J1250 fields, we observe a decrease in LAE surface density within 20 $h^{-1}$ Mpc of the quasar.  As shown in Figure~\ref{fig:models_opaque}, this deficit of LAEs surrounding highly opaque lines of sight is consistent with galaxy UVB and late reionization models, but strongly disfavors the temperature model. We thus demonstrate that the association between high \lya\ opacity and low galaxy density first reported by \citet{becker18} extends to at least two fields.  While \citet{becker18} considered only fluctuating UVB and temperature models, moreover, here we show that the observed opacity-density relation is consistent with models where reionization extends to $z < 6$.

\section{Summary}\label{sec:summary}
We present a selection of Lyman-alpha emitting galaxies (LAEs) using Subaru HSC narrow-band imaging in the fields surrounding two highly opaque quasar sightlines, towards ULAS J0148+0600 ($\tau_\mathrm{eff} \geq7 $) and  SDSS J1250+3130 ($\tau_\mathrm{eff} = 5.7 \pm 0.4$). The observations establish the LAE density expected in the vicinity of two giant \lya\ troughs, which we use to test IGM models that predict a relationship between opacity and density. The results for the J0148 field are an update to those previously reported by \citet{becker18}, here using improved photometric measurements and more stringent LAE selection criteria. 
Observations of the J1250 field are presented here for the first time.

In both fields, we find a deficit of LAEs within 20 $h^{-1}$ Mpc of the quasar sightline. This confirms the results of \citet{becker18} in the J0148 field, and demonstrates in a second field that long, highly opaque \lya\ troughs are associated with underdense regions as traced by LAEs.

These observations provide a direct test of three major types of model that attempt to reproduce the large-scale scatter in \lya\ opacity at $z \simeq $ 5.5--6: fluctuating ultraviolet background models, where the UVB is produced either by galaxies \citep{davies16,daloisio18,nasir20} or quasars \citep{chardin15,chardin17}; the fluctuating temperature model \citep{daloisio15}; and ultra-late reionization models \citep{kulkarni18,nasir20,keating20a,keating20b}. Our results disfavor the temperature model but are consistent with predictions made by galaxy-driven UVB models, in which highly opaque troughs correspond to low-density regions with a suppressed ionizing background.  The results are also consistent with ultra-late reionization models, in which long troughs arise from the last remaining islands of neutral hydrogen, which are also predicted to occur in low-density regions. There is some overlap between these two types of models, as the ultra-late reionization models also include strong UVB fluctuations. The ultra-late reionization model is distinguished by the presence of neutral islands at $z < 6$.

Our results are consistent with a number of recent observations that point towards a late and rapid reionization scenario that has its midpoint at $z\sim7-8$ and ends at $z\leq 6$. A growing body of work is reconsidering the long-standing conclusion that reionization was complete by $z=6$ \citep[e.g.,][]{kulkarni18,keating20a,keating20b,nasir20,choudhury21,qin21}, and therefore discriminating between late reionization and fluctuating UVB models is of great interest. 

This work has focused on fields surrounding highly opaque lines of sight, but further insight may come from fields at the opposite extreme of \lya\ opacity. UVB models predict that highly transmissive sightlines should be associated with galaxy overdensities producing a strong ionizing background \citep{davies18}.  In contrast, late reionization models predict that, in some cases, transmissive sightlines should be associated with low-density regions that have been recently reionized \citep{keating20b}, and may generally arise from a range of overdensities \citep{nasir20}. Establishing the density field surrounding highly transmissive sightlines may therefore prove to be a useful test of these competing models.

\acknowledgments
H.C. is supported by the National Science Foundation Graduate Research Fellowship Program under Grant No. DGE-1326120. G.B. and Y.Z. are supported by the National Science Foundation through grant AST-1615814.

This research is based in part on data collected at Subaru Telescope,
which is operated by the National Astronomical Observatory of Japan.
We are honored and grateful for the opportunity of observing the
Universe from Maunakea, which has cultural, historical and natural
significance in Hawaii.

The Hyper Suprime-Cam (HSC) collaboration includes the astronomical communities of Japan and Taiwan, and Princeton University. The HSC instrumentation and software were developed by the National Astronomical Observatory of Japan (NAOJ), the Kavli Institute for the Physics and Mathematics of the Universe (Kavli IPMU), the University of Tokyo, the High Energy Accelerator Research Organization (KEK), the Academia Sinica Institute for Astronomy and Astrophysics in Taiwan (ASIAA), and Princeton University. Funding was contributed by the FIRST program from Japanese Cabinet Office, the Ministry of Education, Culture, Sports, Science and Technology (MEXT), the Japan Society for the Promotion of Science (JSPS),
Japan Science and Technology Agency (JST), the Toray Science Foundation, NAOJ, Kavli IPMU, KEK, ASIAA, and Princeton University.

Data presented herein were obtained at the W. M. Keck Observatory, which is operated as a scientific partnership among the California Institute of Technology, the University of California and the National Aeronautics and Space Administration. The Observatory was made possible by the generous financial support of the W. M. Keck Foundation.

The Pan-STARRS1 Surveys (PS1) have been made possible through contributions of the Institute for Astronomy, the University of Hawaii, the Pan-STARRS Project Office, the Max-Planck Society and its participating institutes, the Max Planck Institute for Astronomy, Heidelberg and the Max Planck Institute for Extraterrestrial Physics, Garching, The Johns Hopkins University, Durham University, the University of Edinburgh, Queens University Belfast, the Harvard-Smithsonian Center for Astrophysics, the Las Cumbres Observatory Global Telescope Network Incorporated, the National Central University of Taiwan, the Space Telescope Science Institute, the National Aeronautics and Space Administration under Grant No. NNX08AR22G issued through the Planetary Science Division of the NASA Science Mission Directorate, the National Science Foundation under Grant No. AST-1238877, the University of Maryland, and Eotvos Lorand University (ELTE), the Los Alamos National Laboratory, and the Gordon and Betty Moore Foundation.

This paper makes use of software developed for the Large Synoptic Survey Telescope. We thank the LSST Project for making their code available as free software at http://dm.lsst.org.

This research made use of the following additional software: Astropy \citep{astropy2013, astropy2018}, matplotlib \citep{hunter2007}, and scipy\citep{scipy2020}, including numpy \citep{numpy2011}.

The authors wish to recognize and acknowledge the very significant cultural role and reverence that the summit of Mauna Kea has always had within the indigenous Hawaiian community. We are most fortunate to have the opportunity to conduct observations from this mountain. 

\bibliographystyle{apj.bst}
\bibliography{main.bib}

\appendix 
\section{\lya\ Opacity of Quasar Sightlines}
As done by \citet{becker18}, we use our imaging data to estimate $\tau_{\rm{eff}}$ along both quasar lines of sight using the PSF photometry described in section \ref{sec:psf_fitting}. The purpose of these measurements is to check whether our data are consistent with existing spectroscopic limits for these sightlines, and whether it's possible to improve on the existing limits given the depth of our data. For each quasar, we first measure the $NB816$ and $i2$ fluxes. We then convolve each object's spectrum with the $i2$ transmission curve, and scale the spectrum so that its transmission-weighted mean flux over the $i2$ band matches what was measured in the imaging  We use the scaled spectrum to estimate the unabsorbed continuum flux at the narrowband wavelength, and then from the continuum estimate and the photometric narrow-band flux we
calculate the effective opacity as $\tau_{\rm{eff}} = -\rm{ln}(F_{\lambda}^{NB816}/F_{\lambda}^{\rm{cont}})$. These measurements represent an effective opacity over the NB816 wavelength region, which is overlapped by but considerably shorter than the spectroscopically measured regions of both troughs.

In the J0148 sightline, we measure $F_{\lambda}^{NB816}= (2.0\pm 1.8)\times10^{-20}$ erg s$^{-1}$ cm$^{-2}$ \AA$^{-1}$ and $F_{\lambda}^{i2}= (3.2\pm 0.5)\times10^{-18}$ erg s$^{-1}$ cm$^{-2}$ \AA$^{-1}$, and estimate that the unabsorbed continuum is $F_{\lambda}^{\rm{cont}}=1.5\times10^{17}$ erg s$^{-1}$ cm$^{-2}$ \AA$^{-1}$. We therefore measure $\tau_{\rm eff} = 6.63^{+2.5}_{-0.65} (1\sigma)$, or a $2\sigma$ lower limit of $\tau_{\rm eff}\geq 5.59$, which is consistent with the $2\sigma$ lower limit measured by \citet{becker15} of $\tau_{\rm eff}\geq 7.2$ measured over a 50 $h^{-1}$ Mpc section centered at z=5.726.

The J1250 quasar is not detected in our $NB816$ data. As a rough estimate, we adopt the $2\sigma$ upper limit, $F_{\lambda}^{NB816}\leq 4.0\times10^{-20}$ erg s$^{-1}$ cm$^{-2}$ \AA$^{-1}$. We also measure $F_{\lambda}^{i2}= (7.3\pm 0.2)\times10^{-19}$ erg s$^{-1}$ cm$^{-2}$ \AA$^{-1}$, and estimate that the unabsorbed continuum is $F_{\lambda}^{\rm{cont}}=1.0\times10^{17}$ erg s$^{-1}$ cm$^{-2}$ \AA$^{-1}$. We therefore measure $\tau_{\rm eff} \geq 5.52^{+0.69}_{-0.41} (1\sigma)$, or a $2\sigma$ lower limit of $\tau_{\rm eff}\geq 4.83$. This measurement is consistent with that of Zhu et al. (in prep), who find that $\tau_{\rm eff}=5.7\pm0.4$ measured over 81 $h^{-1}$ Mpc centered at z=5.760.

\section{Spectroscopic followup of J0148 LAEs with Keck/DEIMOS}\label{sec:j0148_spec}
\subsection{Observations}\label{sec:deimos_obs}

\begin{figure*}
\includegraphics[width=\textwidth]{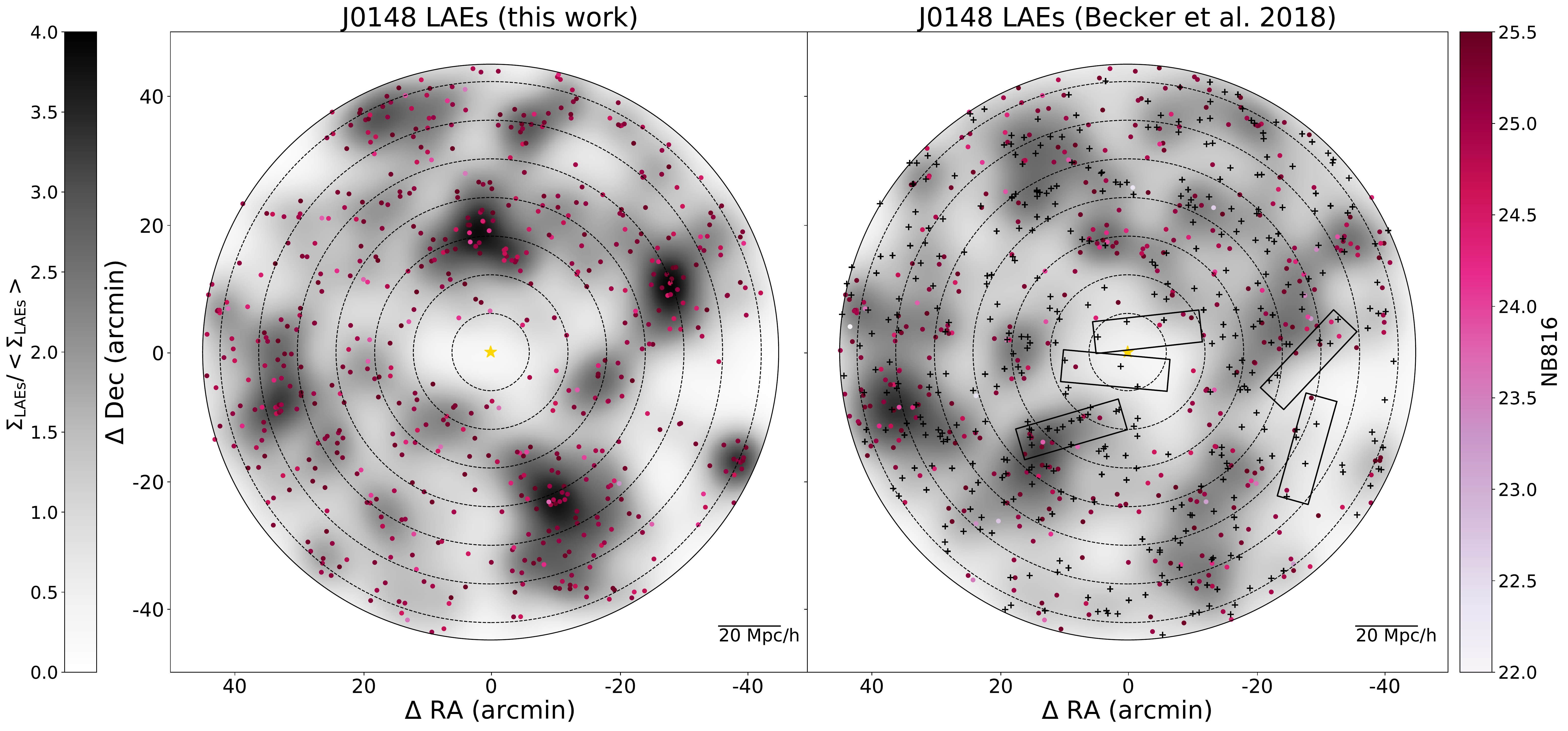}
\caption{Distribution of LAE candidates in the J0148 field as selected in this work (left) and by \citet{becker18} (right). Each field is shown centered on the quasar (gold star). LAE candidates with NB816<25.5 are shown with a color that indicates their narrowband magnitude in their respective catalog. In the right panel, LAEs selected in \citet{becker18} with NB816>25.5 (fainter than our selection limit) are shown with black crosses. The surface density at each position is calculated by kernel density estimation using a Gaussian kernel, and is normalized by the mean surface density measured across the entire field.  Concentric, dotted circles are shown in increments of 10 Mpc h$^{-1}$ projected distance from the quasar. The black rectangles in the right panel indicate the pointings of DEIMOS slitmasks used for spectroscopic followup. The solid circle marks the edge of the field of view, 45\arcmin\ from the quasar.}
\label{fig:j0148_comparison_laes}
\end{figure*}

\LongTables
\begin{center}
\begin{deluxetable}{lcccc}
\tablewidth{0pc}
\tablecaption{Summary of Keck/DEIMOS observations \label{tab:spec_summary}}
\tabletypesize{\scriptsize}
\tablehead{\colhead{Date} & \colhead{Mask}$^a$ & \colhead{Description} & \colhead{$t_{exp}$ (h)} & \colhead{Seeing$^b$}}
\startdata
11/28/18 & 1 & Central & 2 & 0.74\arcsec\\
11/28/18 & 2 & Central & 2 & 0.73\arcsec\\
11/29/18 & 3 & High Density & 2 & 0.83\arcsec \\
11/29/18 & 4 & High Density & 2 & 0.65\arcsec\\
11/28-11/29 & 5 & Low Density & 1.2 & 0.78\arcsec
\enddata
\tablenotetext{a}{The first priority for mask placement was to maximize the number of LAE candidates observed within 20 h$^{-1}$ Mpc of the quasar. Masks were also placed to cover other high- and low-density regions of the field.}
\tablenotetext{b}{Median seeing measured from Gaussian fits to the profiles of stars on each mask.}
\end{deluxetable}
\end{center}

~\label{spec_summary}

In addition to the imaging data discussed in Section~\ref{sec:obs}, we obtained spectra of 46 LAE candidates in the J0148 field taken with the DEIMOS spectrograph \citep{faber2003} on the Keck II telescope in November 2018. Targets were selected from the catalog of LAE candidates published in \citet{becker18}, as spectroscopic followup was carried out prior to the creation of the catalog presented in this work. We prioritized objects in the underdense region at the center of the field of view, a second low-density region at the west edge of the field, and a high-density region. In total we used 5 masks, which we designed using DSIMULATOR (Figure \ref{fig:j0148_comparison_laes}). The observations, which are summarized in Table~\ref{spec_summary}, were made in multi-slit mode using the OG550 filter and the 600-line grating. Each individual target was placed in a 1\arcsec\ slit, and all slits were tilted five degrees relative to the position angle of the mask in order to better sample the sky lines for sky subtraction.

We reduced the raw spectra with a custom IDL pipeline that includes optimal sky subtraction \citep{kelson2003}. Individual exposures were then combined onto a two-dimensional grid rectified with nearest neighbor resampling, in which each frame’s individual pixels are assigned to the pixel in the combined frame that it most closely matches in position and wavelength. Rectifying the spectra in this way ensures that pixels in the combined frame remains uncorrelated. Finally, we corrected the spectra for atmospheric absorption, and flux calibrated using standard stars.

\subsection{Results}
\begin{figure}
\begin{center}
\includegraphics[width=0.45\textwidth]{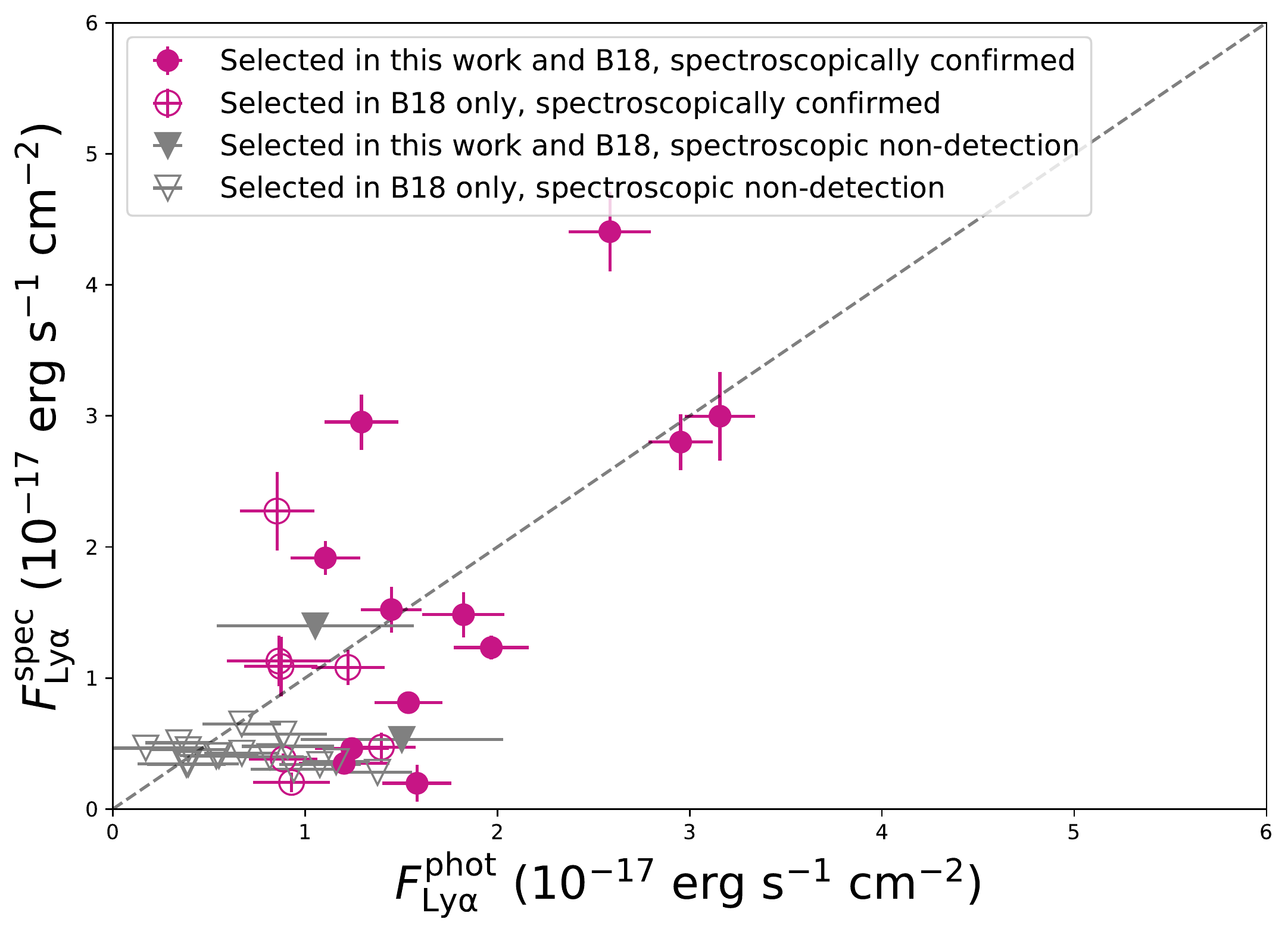}
\caption{Photometric and spectroscopic fluxes for all credible LAE candidates. The following objects are considered credible: all spectroscopically confirmed objects, spectroscopic non-detections that were selected photometrically in this work, and non-detections that were selected only by \citet{becker18} that also passed a secondary visual inspection to remove clearly spurious sources.
Spectroscopically confirmed LAEs are shown with circles, and spectroscopic non-detections are shown with triangles. LAEs that meet the photometric criteria outlined in ~\ref{sec:lae_selection} are shown with filled markers; LAEs that fail one or more photometric criteria are shown with empty markers. For spectroscopic non-detections, the reported flux is a $1\sigma$ upper limit.}
\label{fig:lya_luminosity}
\end{center}
\end{figure}
\begin{figure*}
\includegraphics[width=\textwidth]{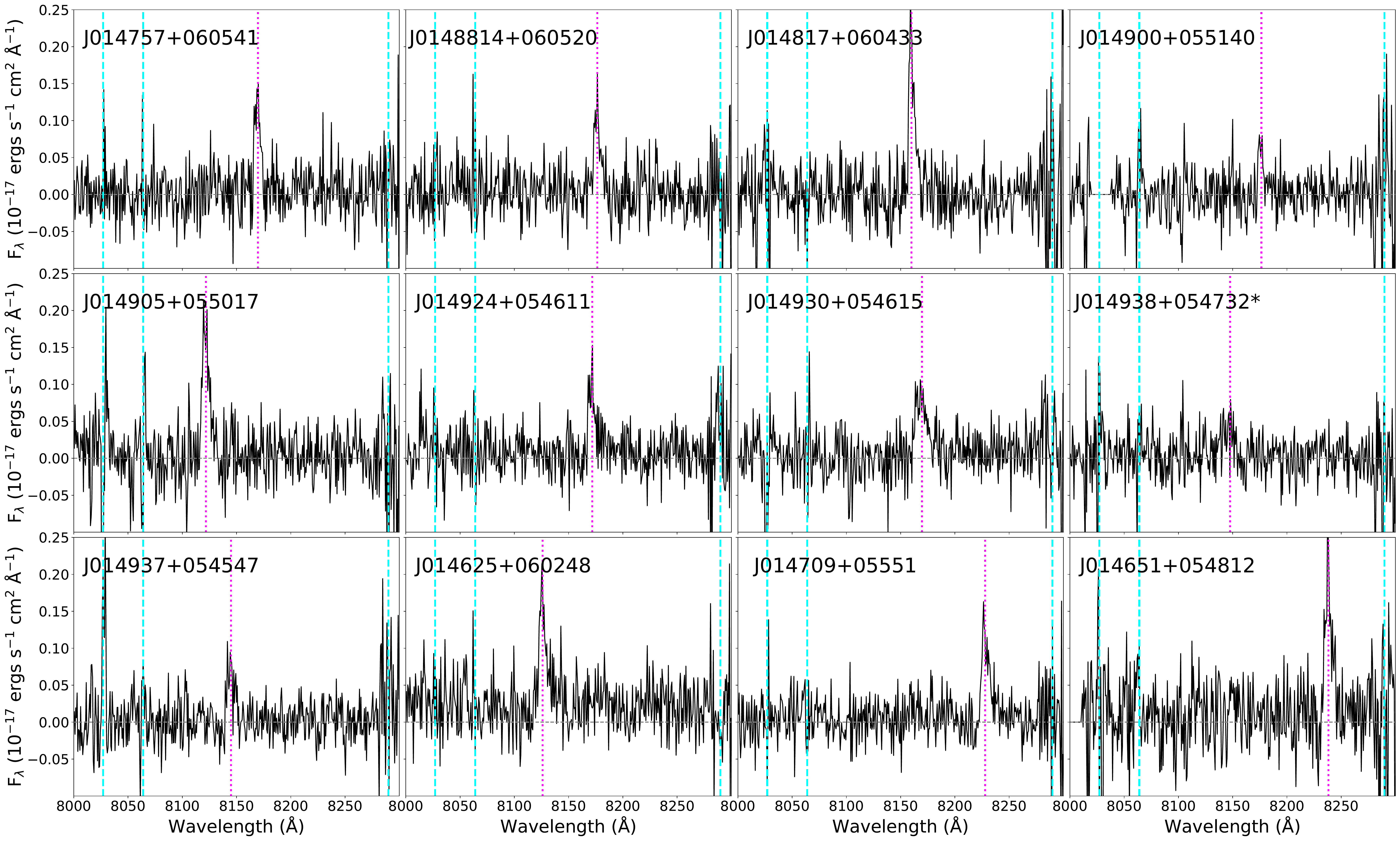}
\caption{Spectra for each of the spectroscopically confirmed LAEs that meet the photometric criteria outlined in ~\ref{sec:lae_selection}. The dashed cyan lines indicate skyline residuals, while the dotted pink line indicates the flux-weighted mean wavelength of the emission line, which is used to calculate the spectroscopic redshift. J0149938+054732 (marked with an asterisk) is a marginal detection with 1.4$\sigma$ confidence.}
\label{fig:spec_laes_1D}
\end{figure*}

\begin{figure*}
\includegraphics[width=\textwidth]{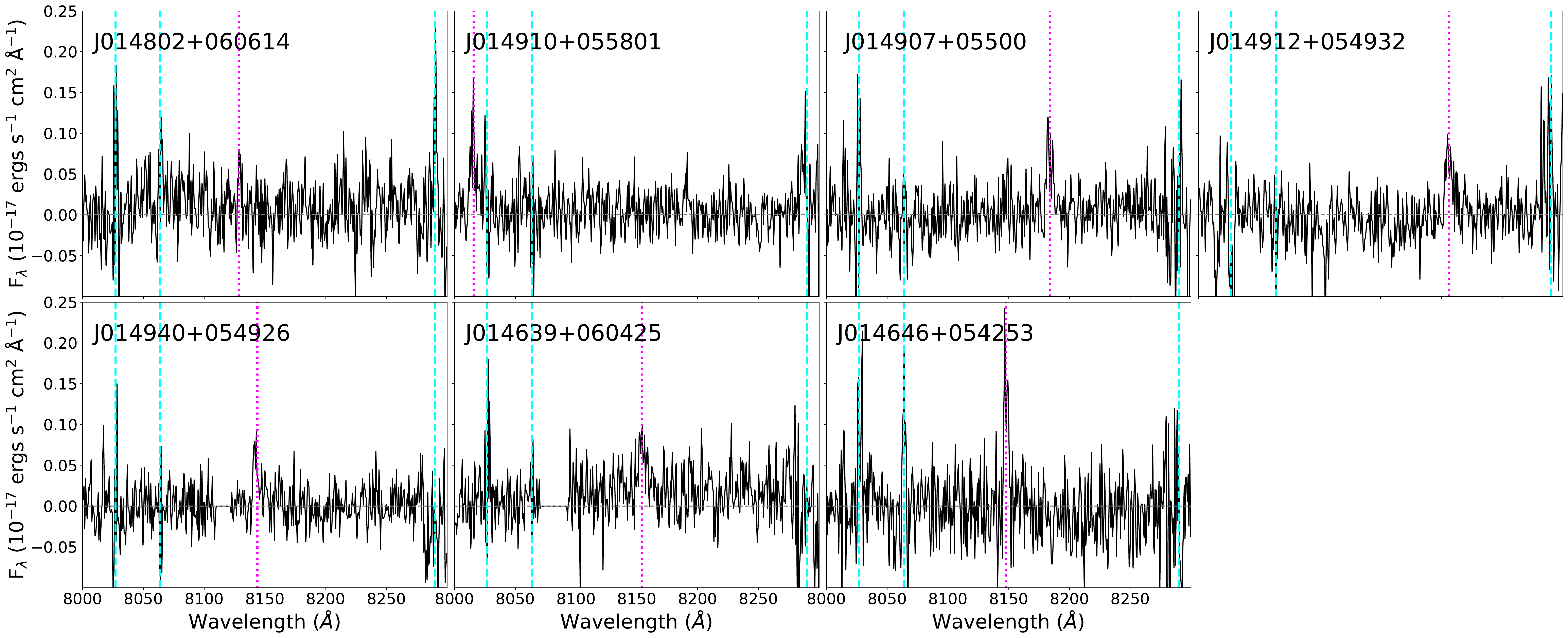}
\caption{Spectra for each of the spectroscopically confirmed LAEs that were selected in \citet{becker18} but were not selected by the photometric criteria outlined in ~\ref{sec:lae_selection}. The dashed cyan lines indicate skyline residuals and the dotted pink line indicates the flux-weighted mean wavelength of the emission line, which is used to calculate the spectroscopic redshift. }
\label{fig:spec_laes_1D-2}
\end{figure*}

\LongTables
\begin{center}
\begin{deluxetable*}{ccccccccc}
\tablewidth{1.75\textwidth}
\tablecaption{Summary of \hmc{all} spectroscopically confirmed LAEs in the J0148 field \label{tab:spec_laes}}
\tabletypesize{\scriptsize}
\tablehead{\colhead{ID} & \colhead{RA (J200)} & \colhead{Dec (J1200)} &\colhead{z$_{\rm{spec}}$} & \colhead{$m_{NB816}^a$ (mag)} & \colhead{F$_{phot}\cdot10^{17}$} & \colhead{F$_{spec}\cdot10^{17}$} & \colhead{Selected?$^c$}}
\startdata
J014757+060541&01h47m57.824s & +06d05m41.87s&5.72 & 25.12 & 1.54 $\pm$ 0.18 & 0.81 $\pm$ 0.08 & Y \\
J014802+060614&01d48m02.906s & +06d06m14.46s&5.69 & 25.67 & 0.93 $\pm$ 0.2 & 0.2 $\pm$ 0.08 & N \\
J0148814+060520&01h48m14.932s&+06d05m20.174s&5.73 & 25.19 & 1.45 $\pm$ 0.16 & 1.52 $\pm$ 0.17 & Y \\
J014817+060433&01h48m17.678s&+06d04m33.38s&5.71 & 24.42 & 2.95 $\pm$ 0.17 & 2.8 $\pm$ 0.21 & Y \\
J014910+055801&01h49m10.161s&+05d58m01.87s&5.59 & 25.23 & 1.4 $\pm$ 0.18 & 0.47 $\pm$ 0.11 &N \\
J014900+055140&01h49m00.853s&+05d51m40.93s&5.73 & 25.39 & 1.2 $\pm$ 0.23 & 0.35 $\pm$ 0.07 & Y \\ 
J014905+055017&01h49m05.023s&+05d50m17.85s& 5.68 & 24.56 & 2.58 $\pm$ 0.21 & 4.41 $\pm$ 0.3 & Y \\
J014907+05500&01h49m07.708s&+05d50m01.74s&5.73 & 25.75 & 0.86 $\pm$ 0.27 & 1.13 $\pm$ 0.19 & N \\ 
J014912+054932&01h49m12.801s&+05d49m32.61s&5.75 & 25.37 & 1.22 $\pm$ 0.19 & 1.08 $\pm$ 0.14 & N \\ 
J014924+054611&01h49m24.820s&+05d46m11.47s&5.72 & 24.94 & 1.82 $\pm$ 0.21 & 1.48 $\pm$ 0.17 & Y \\
J014930+054615&01h49m30.632s&+05d46m15.74s&5.72 & 24.34 & 3.16 $\pm$ 0.18 & 3.0 $\pm$ 0.34 & Y \\
J014940+054926&01h49m40.087s&+05d49m26.15s&5.70 & 25.72 & 0.89 $\pm$ 0.18 & 0.38 $\pm$ 0.04 & N \\
J014938+054732$^b$ & 01h49m38.827s & +05d47m32.37s & 5.70 & 25.09 & 1.58 $\pm$ 0.18 & 0.2 $\pm$ 0.14 & Y \\
J014937+054547&01h49m37.493s&+05d45m47.48s&5.70 & 25.36 & 1.24 $\pm$ 0.19 & 0.46 $\pm$ 0.05 &Y \\
J014625+060248&01h46m25.501s&+06d02m48.51s&5.69 & 24.86 & 1.97 $\pm$ 0.2 & 1.23 $\pm$ 0.09 &Y\\ 
J014639+060425&01h46m39.395s&+06d04m25.49s&5.71 & 25.76 & 0.85 $\pm$ 0.19 & 2.27 $\pm$ 0.3 & N \\
J014709+05551&01h47m09.103s&+05d55m51.99s&5.77 & 25.48 & 1.11 $\pm$ 0.18 & 1.92 $\pm$ 0.13 & Y \\
J014651+054812&01h46m51.818s&+05d48m12.57s&5.78 & 25.31 & 1.29 $\pm$ 0.19 & 2.95 $\pm$ 0.21 & Y \\
J014646+054253&01h46m46.656s&+05d42m53.31s&5.70 & 25.74 & 0.87 $\pm$ 0.19 & 1.09 $\pm$ 0.23 & N \enddata
\tablenotetext{a}{Photometric measurement}
\tablenotetext{b}{Marginal detection}
\tablenotetext{c}{Indicates whether this object met the updated selection criteria described in Section~\ref{sec:methods}.}
\end{deluxetable*}
\end{center}
\label{spec_laes}

Emission lines were identified by visual inspection of the 2D spectra. To be spectroscopically confirmed, a LAE candidate was required to have a single emission line in the \lya\ region, and no emission lines elsewhere in the trace. For each spectroscopically confirmed LAE, we determine the spectroscopic redshift from the flux-weighted mean wavelength of the emission line, which is calculated over a 20 \AA\ window centered on the visually estimated line center. This window was chosen to be wide enough to cover any of the emission lines in our sample, but not so wide as to include unwanted skyline noise. We also measure the \lya\ flux by integrating the spectrum over a wavelength region that includes the entire emission line; this region is customized for each object, but is typically $\sim15$ \AA. Table~\ref{tab:spec_laes} summarizes the properties of all spectroscopically confirmed LAEs. We compare the photometric and spectroscopic \lya\ fluxes for all credible LAEs in Figure~\ref{fig:lya_luminosity}, which includes spectroscopically confirmed objects, spectroscopic non-detections that were selected photometrically in this work, and non-detections that were selected only by \citet{becker18} that also passed a secondary visual inspection to remove clearly spurious sources. Figure~\ref{fig:lya_luminosity} demonstrates a reasonable agreement between the photometric and spectroscopic measurements, including for the spectroscopic non-detections, which tend to be the faintest objects in the sample.

 Among the 46 LAE candidates from \citet{becker18} targeted for spectroscopic follow-up, 14 were also selected as LAEs in this work using the updated PSF photometry and the new LAE selection criteria. Of those 14, 11 were spectroscopically confirmed at $\geq4\sigma$ confidence, and one was marginally detected at 1.4$\sigma$. Figure~\ref{fig:spec_laes_1D} shows 1D spectra for these PSF-selected LAEs. The dashed cyan lines indicate skyline residuals, and the dotted pink line line indicates the flux-weighted mean wavelength of the emission line. 

The remaining 32 objects targeted for spectroscopic followup were selected as LAEs only by \citet{becker18}. Of these, seven are spectroscopically confirmed LAEs, and their 1D spectra are shown in Figure~\ref{fig:spec_laes_1D-2}. These seven fell just outside our new selection criteria using the updated PSF photometry; four had narrowband
$4.5 < S/N < 5$, and one had $S/N=3.2$.
The remaining two are detected in the $r2$ band at 2.3$\sigma$, which is slightly higher than our $r2$ cuts allow. The other 25 objects failed our updated selection criteria by wider margins. Their spectroscopic non-detections are attributed to the issues with CModel fluxes described in Section~\ref{sec:psf_fitting}, with the exception of one object, which was a low-redshift contaminant displaying a clear [OIII] emission line.

In summary, we find a high spectroscopic confirmation rate (11 plus one marginal detection out of 14) among candidates selected using our updated photometry and selection criteria. The two non-detected objects of the photometrically selected group were generally fainter than their detected counterparts, with a $1\sigma$ upper limit on their flux being consistent with the photometric measurement, and showed no sign of being low-redshift contaminants. We note that all of the objects followed up spectroscopically were also selected as LAEs by \citet{becker18}, so these 14 candidates do not represent an unbiased random sample from the new photometric catalog. Nevertheless, the high confirmation rate among the PSF-selected candidates gives us confidence that the photometric selection methods described in Section \ref{sec:lae_selection} should yield a high-fidelity sample of LAEs.

\section{Comparison to Becker 2018 LAE catalog}\label{sec:j0148_comparison}
Here we compare the LAE catalog presented in this work, using updated photometry and selection criteria as described in Sections~\ref{sec:lae_selection} and~\ref{sec:psf_fitting}, to that published in \citet{becker18}.

\LongTables
\begin{center}
\begin{deluxetable}{lcc}
\tablewidth{0pc}
\tablecaption{Comparison of LAE selections in \citet{becker18} and this work \label{tab:j0148_comparison}}
\tabletypesize{\scriptsize}
\tablehead{\colhead{} & \colhead{This Work } & {\citet{becker18}}}

Objects selected with $NB816\leq26.0$ & 784$^{a}$ & 806\\
Objects selected with $NB816\leq25.5$ & 641 & 398$^b$\\ 
\hline 
& \multicolumn{2}{c}{Both Works} \\
\hline 
Catalog overlap with $NB816\leq26.0^c$ & \multicolumn{2}{c}{366} \\
Catalog overlap with $NB816\leq25.5^d$ & \multicolumn{2}{c}{236} \\ 
Catalog overlap with published $NB816$ limits$^e$ & \multicolumn{2}{c}{321} 
\enddata
\tablenotetext{a}{This work uses a brighter magnitude limit than \citet{becker18} ($NB816\leq25.5$). The number of objects selected with the fainter limit is included only for comparison. }
\tablenotetext{b}{Likewise, \citet{becker18} use $NB816\leq26.0$. The subset of this catalog that satisfies the brighter magnitude limit is included here for comparison.}
\tablenotetext{c}{The number of LAEs appearing in both catalogs that meet the fainter magnitude requiprement. $NB816\leq 26.0$ (as in \citealt{becker18})}
\tablenotetext{d}{The number of LAEs appearing in both catalogs that meet the brighter magnitude requirement, $NB816\leq 25.5$ (as in this work)}
\tablenotetext{e}{The number of LAEs appearing in both catalogs as is, using $NB816\leq25.5$ for this work and $NB816\leq26.0$ for \citet{becker18} (as published)}

\end{deluxetable}
\end{center}
~\label{j0148_comparison}

\begin{figure}
\begin{center}
\includegraphics[width=0.45\textwidth]{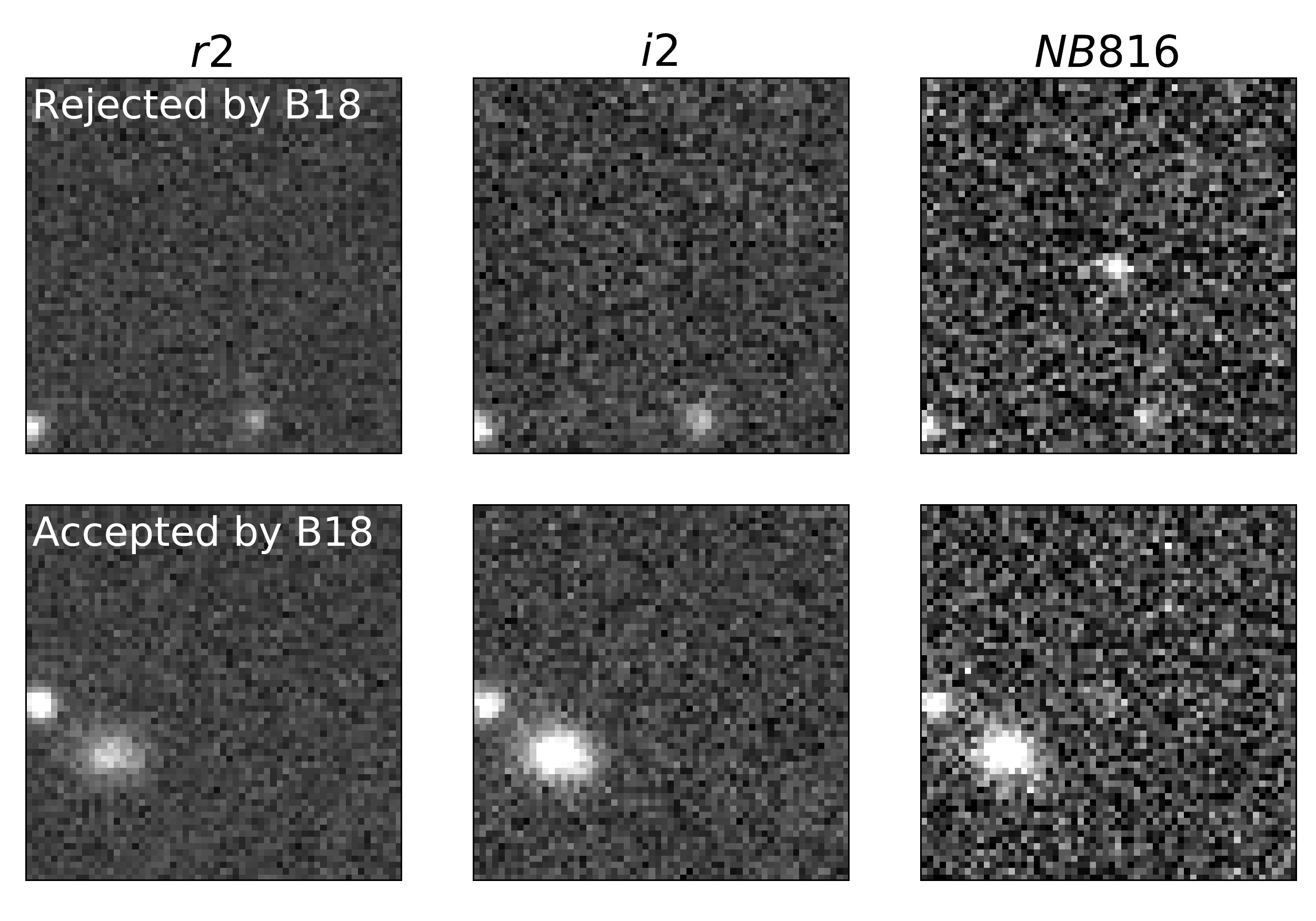}
\caption{Examples of objects rejected (top row) and accepted (bottom row) as LAE candidates by \citet{becker18} based on spurious CModel photometry. Each cutout is 10\arcsec\ on each side and centered on the object position. The rejected object is detected in the CModel photometry at $10\sigma$ in the narrowband, $18\sigma$ in $i2$, and $26\sigma$ in $r2$ - a clear case of artificially high broadband photometry. This object is selected as an LAE in this work using the photometry and selection criteria outlined in Section~\ref{sec:methods}. The accepted object is undetected in the broadbands, but is detected using the CModel photometry at $7.5\sigma$ in the narrowband, compared to $3.0\sigma$ using our PSF photometry. }
\label{fig:example_failures}
\end{center}
\end{figure}

In this work, we identify \numLAESa\ LAE candidates in the J0148 field, compared to 806 LAEs presented in \citet{becker18}. Of the objects selected by \citet{becker18}, 398 had $NB816<25.5$ as required in this work, and 236 of those objects ($\sim$60\%) are selected using the selection criteria and photometric measurements outlined in Section~\ref{sec:methods}.  We estimate that 15-20\% of the objects selected by \citet{becker18} with $NB816<25.5$ were affected by the systematic CModel flux effects described in Section~\ref{sec:methods}.  We show examples of objects wrongly rejected and accepted due to these issues in Figure~\ref{fig:example_failures}. Each cutout image is 10\arcsec\ on each side and centered on the object position. The wrongly rejected object was rejected based on artificially high broadband fluxes, while the wrongly accepted object had inflated $NB816$ flux.  The remaining 20-25\% of the \citet{becker18} objects with $NB816<25.5$ missing from our sample are within $1\sigma$ errors of meeting our selection criteria. Given that our catalog is $\sim$ 50\% complete at the faintest magnitudes, it is not unexpected that some objects will not be selected due to photometric scattering.

Table \ref{tab:j0148_comparison} summarizes the number of LAEs selected in both catalogs. Because the two catalogs use different narrowband magnitude limits, $NB816\leq 25.5$ in this work and $NB816\leq 26.0$ in \citet{becker18}, we provide the number of objects selected in each catalog using both limits. We emphasize that this work only makes use of $NB816\leq25.5$ objects for our analysis; the fainter magnitude limit is provided only for comparison. Table \ref{tab:j0148_comparison} also summarizes the number of LAEs that are common to both catalogs using both magnitude limits, as well as the number of objects common to the catalogs as is (using $NB816\leq25.5$ for the objects selected in this work, and $NB816\leq26.0$ for \citet{becker18}, as published).


Figure~\ref{fig:j0148_comparison_laes} shows the distribution of LAE candidates in the J0148 field, as presented in this work (left) and in \citet{becker18} (center). Each LAE is color-coded according to the $NB816$ magnitude in its respective catalog. This work has a shallower narrowband magnitude limit than \citet{becker18}; we have therefore shown LAEs that fall in the $25.5 \leq\ NB816 \leq 26.0$ bin from the \citet{becker18} catalog with black crosses, as they are fainter than our selection criteria allow. The quasar (yellow star) is centered in each panel, and the dotted concentric circles show increments of 10 $h^{-1}$ Mpc. The solid outer circle marks the edge of the field of view, 45\arcmin\ from the quasar. LAE candidates are shown plotted over a surface density map, which we create by kernel density estimation over a regular grid of 0.24\arcmin\ pixels using a Gaussian kernel of bandwidth 1.6\arcmin. The surface density map is normalized by the mean surface density of the field. While the exact membership is varied between the two catalogs, both show similar large-scale structures.

\begin{figure}
\begin{center}
\includegraphics[width=0.45\textwidth]{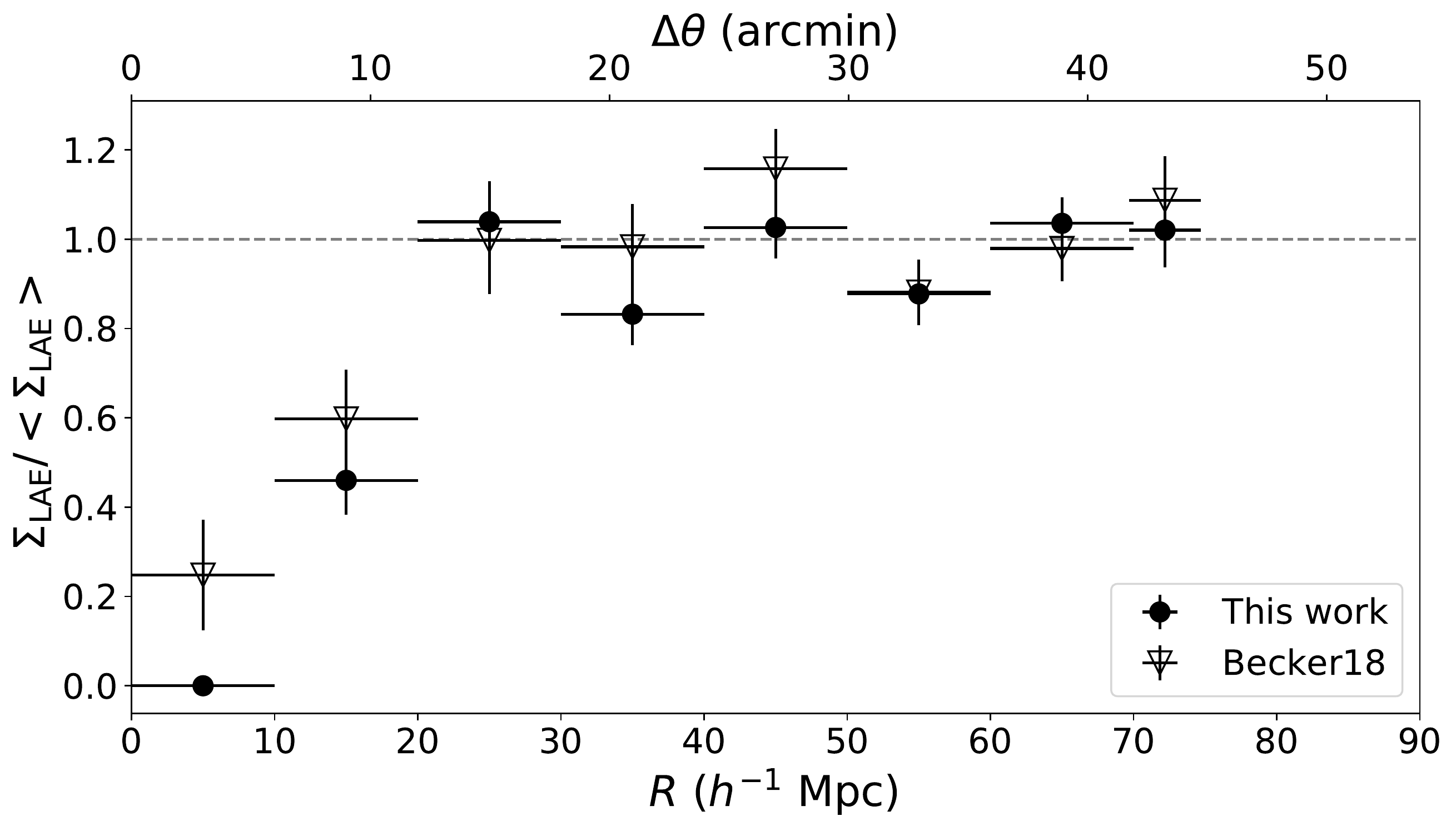}
\caption{Surface density of completeness-corrected LAEs in the J0148 field, as selected in this work (filled circles) and by \citet{becker18} (open triangles). The surface density is measured in annuli of width 10 $h^{-1}$\, Mpc for all except the outermost bin, which has a width of 4.5 $h^{-1}$\,Mpc. All surface densities are normalized by the mean value in the respective work, calculated over $15\arcmin \leq \theta \leq 40\arcmin$. Horizontal error bars show the width of the annuli, and vertical error bars are 68\% Poisson intervals.}
\label{fig:j0148_comparison_rad}
\end{center}
\end{figure}

Figure~\ref{fig:j0148_comparison_rad} shows the surface density as a function of radial distance from the quasar in the J0148 field, as measured here (circles) and by \citet{becker18} (triangles). The surface densities are measured in 10 $h^{-1}$ Mpc annuli for all except the outermost bin, which is 4.5 $h^{-1}$ Mpc, and normalized by the mean surface density, which is measured over $15\arcmin \leq \theta \leq 40\arcmin$. We note that, in addition to the changes to fluxing and LAE selection criteria, the completeness corrections used in this work (see Section~\ref{sec:results}) are different than those used by \citet{becker18}. However, in most radial bins the surface density measurements are consistent within the 1$\sigma$ errors.

To summarize, the results in the J0148 field are largely unchanged between this work and \citet{becker18}. Approximately 50\% of the LAEs selected in this work are also selected by \citet{becker18}, and, outside of the photometry issues described in Section~\ref{sec:psf_fitting}, the variations are as expected given that each catalog is $\sim50$ \% complete in its faintest magnitude bin. The two catalogs trace similar large-scale structures (see Figure~\ref{fig:j0148_comparison_laes}), most notably both displaying the $\sim20$ $h^{-1}$ Mpc void in the center of the field, along the quasar line of sight.

\section{Completeness Corrections}

\begin{figure}
\begin{center}
\includegraphics[width=0.8\textwidth]{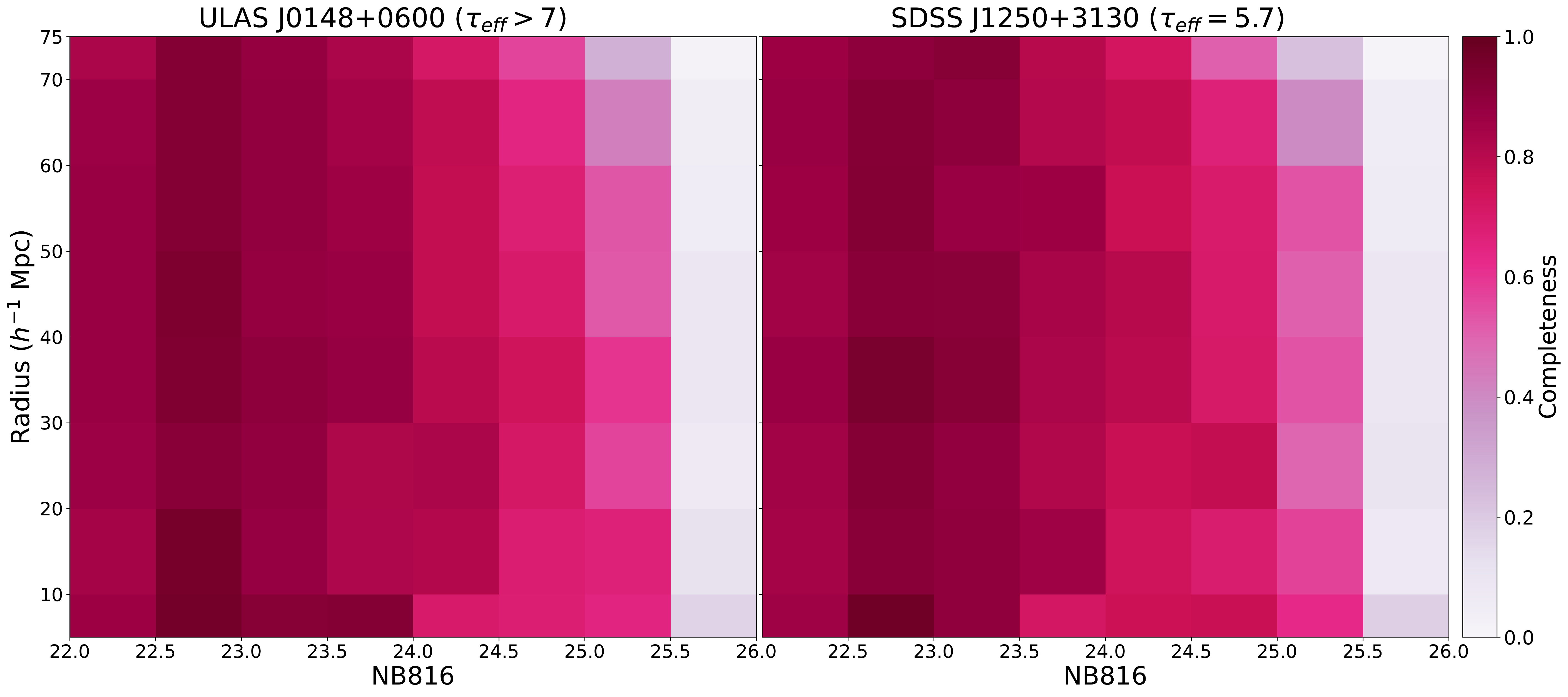}
\caption{Completeness measured the J0148 field (left) and J1250 field (right) as a function of NB816 magnitude and distance to the quasar. The completeness is based on the detection rate of artificial LAEs injected in each field, and is given by fraction of artificial LAEs detected in each radius and magnitude bin.}
\label{fig:completeness_maps}
\end{center}
\end{figure}

Figure~\ref{fig:completeness_maps} shows the completeness measured in both fields as a function of distance from the quasar and NB816 magnitude. We determine the completeness by injecting a catalog of artificial LAE candidates across each field and then applying the LAE selection criteria described in Section~\ref{sec:methods}. We bin the artificial LAEs by magnitude and distance from the quasar.  The completeness is then computed as the fraction of artificial LAEs detected in each bin. The observations are binned in the same way and corrected by the reciprocal of the completeness in each bin. 

\end{document}